\documentclass[twocolumn,aps,prd,amssymb,eqsecnum,nofootinbib,superscriptaddress]{revtex4}
\usepackage{epsfig}
\usepackage{amsmath}
\usepackage{amssymb}
\usepackage{color}

\def\en{\end{equation}}
\def\ena{\end{eqnarray}}

\def\rmd{{\rm d}}

\begin{document}
\title{Forced motion near black holes}

\author{Jonathan R. Gair}
\affiliation{Institute of Astronomy, Madingley Road, Cambridge CB3 0HA, UK}
\author{\'{E}anna \'{E}. Flanagan}\affiliation{Center for Radiophysics and Space Research, Cornell University, Ithaca, NY 14853, USA}
\author{Steve Drasco}
\affiliation{Albert-Einstein-Institut, Max-Planck-Institut f\"ur Gravitationsphysik, D-14476 Golm, Germany}\affiliation{Theoretical Astrophysics, California Institute of Technology, Pasadena, CA 91125, USA}
\affiliation{California Polytechnic State University, San Luis Obispo, CA 93405,
 USA}
 \author{Tanja Hinderer}
\affiliation{Theoretical Astrophysics, California Institute of Technology, Pasadena, CA 91125, USA}
\author{Stanislav Babak}
\affiliation{Albert-Einstein-Institut, Max-Planck-Institut f\"ur Gravitationsphysik, D-14476 Golm, Germany}

%
%
\newcount\hh
\newcount\mm
\mm=\time
\hh=\time
\divide\hh by 60
\divide\mm by 60
\multiply\mm by 60
\mm=-\mm
\advance\mm by \time
\def\hhmm{\number\hh:\ifnum\mm<10{}0\fi\number\mm}


\date{draft of December 17, 2010; printed \today{} at \hhmm}

\newcommand{\be}{\begin{equation}}
\newcommand{\ee}{\end{equation}}
\newcommand{\bea}{\begin{eqnarray}}
\newcommand{\eea}{\end{eqnarray}}
\newcommand{\bes}{\begin{subequations}}
\newcommand{\ees}{\end{subequations}}


\begin{abstract}
We present two methods for integrating forced geodesic equations in the Kerr spacetime.
The methods can accommodate arbitrary forces.  As a test case, we compute inspirals caused by a simple drag force,
mimicking motion in the presence of gas.  We verify that both methods give the same results for this simple force.
We find that drag generally causes eccentricity to increase throughout the inspiral.
This is a relativistic effect qualitatively opposite to what is seen in gravitational-radiation-driven inspirals,
and similar to what others have observed in hydrodynamic simulations of gaseous binaries.
We provide an analytic explanation by deriving the leading order relativistic correction to the Newtonian dynamics.
If observed, an increasing eccentricity would thus provide clear evidence that the inspiral was occurring in a nonvacuum environment.

Our two methods are especially useful for evolving orbits in the adiabatic regime.
Both use the method of osculating orbits, in which each point on
the orbit is characterized by the parameters of the geodesic with the same instantaneous position and velocity.
Both methods describe the orbit in terms of the geodesic
energy, axial angular momentum, Carter constant, azimuthal phase, and two angular variables that increase monotonically and are relativistic generalizations of the eccentric anomaly. The two methods differ in their treatment of the orbital phases and the representation of the force. In the first method, the geodesic phase and phase constant are evolved together as a single orbital phase parameter, and the force is expressed in terms of its components on the Kinnersley orthonormal tetrad. In the second method, the phase constants of the geodesic motion are evolved separately and the force is expressed in terms of its Boyer-Lindquist components. This second approach is a direct generalization of earlier work by Pound and Poisson~\cite{pound08} for planar forces in a Schwarzschild background.
\end{abstract}

\maketitle
\def\bfpsi{\mbox{\boldmath $\psi$}}
\def\bfchi{\mbox{\boldmath $\chi$}}
\def\bfcalJ{\mbox{\boldmath ${\cal J}$}}

\section{Introduction}
\label{intro}
The two-body problem in relativity when one of the bodies is much more massive than the other is of great interest both theoretically and astrophysically. In this limit, the orbit of the smaller body is approximately geodesic on short time scales. Deviations from the geodesic trajectory arise from the back-reaction on the orbit of the spacetime perturbation created by the object, but can also arise from external factors such as gravitational interactions with other bodies, gaseous material in the spacetime and so forth. In all these situations, the orbit can be described as a geodesic acted on by a perturbing force, which is in general small. In this article, we describe techniques for integrating the Kerr geodesic equations in the presence of an arbitrary forcing term, which can be applied to any of these problems.

For the back-reaction on the orbit, the perturbing force, called the self-force, is of the order of the mass ratio $\mu/M$ and it can be computed by a perturbation expansion in this small parameter. Computing the linearized metric perturbation sourced by the compact object and hence the self-force is not an easy task and it has taken more than a decade to solve this problem for a nonspinning compact object moving in a Schwarzschild background \cite{Vega:2009qb, Barack:2010tm, Dolan:2010mt, Barack:2009ux}. The conventional approach treats the compact object as a test mass which leads to a divergence of the field at the position of the particle and this must be dealt with using a regularization procedure. The extension to Kerr orbits is underway. The techniques described in this paper will be a useful tool in the future for constructing trajectories evolving under gravitational radiation-reaction.

The problem of the motion of two bodies with very different masses is relevant for present and future gravitational wave detectors. Systems with mass ratios of 1:100 (intermediate-mass-ratio inspirals) could be detected by the advanced generation of ground-based detectors that are currently under construction~\cite{LIGOIMRI}. The proposed space-based detector LISA~\cite{SRD} is expected to detect $\sim 10-100$ extreme-mass-ratio inspiral (EMRI) events per year~\cite{gairEMRIrate}. These result from the capture of a compact stellar-mass object (a white dwarf, neutron star or black hole) by a massive black hole (MBH) from a surrounding cusp of stars in a galactic nucleus. The captured object generates a large number of gravitational wave cycles while it is orbiting in the strong field of the MBH, which makes these very good sources to use as probes of strong-field gravity~\cite{EMRIrev}. For both of these classes of source, techniques for evolving the orbit under the influence of both gravitational back-reaction and other perturbing forces are essential for constructing accurate waveform templates and for understanding how external perturbations can leave an imprint on the inspiral trajectory

 We present two implementations that can be used to integrate geodesic motion in a Kerr background with an external force.  We use the method of osculating elements extending previous work \cite{pound08} for Schwarzschild orbits to the Kerr background. The problem of motion under a small perturbation is well studied in celestial mechanics and is regularly applied to model the motion of satellites and small planets.  A geodesic in Newtonian mechanics, or relativity, is uniquely characterized either by the three components of the particle position vector, ${\bf r}$, and the three components of the particle velocity, ${\bf \dot{r}}$, at any time or by six orbital constants (three orbital constants of the motion and three initial phases).
 There is a one-to-one correspondence between the two
 characterizations. This means that {\it any} trajectory can be
 instantaneously identified with a geodesic that has the same values
 of ${\bf r}$ and ${\bf \dot{r}}$.  Of course, at two different
 instances of time, the geodesics will differ, but one can smoothly
 evolve the geodesic parameters to reproduce any nongeodesic
 trajectory. There are several approaches to do so and we describe
 these in the next subsection.

 \subsection{Osculating Elements or variation of constants}
As mentioned above we can describe a bound stable geodesic by six
parameters, which we denote by $I$.  In the nonrelativistic case
these parameters are simply $I = ( {\bf r}, {\dot {\bf r}})$, while
for geodesic motion in Kerr we can take $I = \{E, L_z, Q, \psi_0,
\phi_0, \chi_0\}$.  Here $E$ is the energy, $L_z$ the azimuthal
angular momentum, $Q$ is the Carter constant, and the remaining phases
are defined in Sec.\ \ref{Kerr} below.

At each instant we can therefore identify the true
trajectory with a corresponding geodesic such that  ${\bf r}$ and ${\bf\dot{r}}$ are the same. This
imposes a particular choice of parameters, $I$, at each instance of time, and the whole
trajectory is thus described by a sequence in the geodesic phase space, e.g.,  $I(t) = \{E(t), L_z(t), \iota(t), \psi_0(t), \phi_0(t), \chi_0(t)\}$. These are referred to as the {\it osculating orbital elements} at the {\it osculation epoch} $t$ \cite{bookPert}. Another name for this
approach used in the Hamiltonian description is a {\it variation of constants}.
We preserve the form of the equations of motion for a geodesic but slowly vary what used to be constants of motion in the unperturbed case. There are well known techniques for tackling such problems which are widely used in Newtonian celestial mechanics and can be extended to the relativistic regime. This was demonstrated by Pound and Poisson~\cite{pound08} for the trajectory of a particle in a Schwarzschild background under the action of (post-Newtonian) radiation reaction.

When we have a perturbed system of the form
\be
{\bf \ddot{r}} = {\bf f}_{\rm geo} + \delta{\bf f}~,
\en
we can describe the perturbed trajectory using the osculating elements referred to the orbits of the geodesic system ${\bf \ddot{r}} = {\bf f}_{\rm geo}$. From the chain rule, any one of the osculating elements evolves as
\be
\dot{I} = \nabla_rI\cdot{\bf\dot{r}} + \nabla_v I\cdot{\bf\ddot{r}}~,
\en
in which the subscripts $r$ and $v$ denote derivatives with respect to the orbital position and velocity respectively. In the absence of the perturbing force, each osculating element is constant, so $\dot{I}=\nabla_rI\cdot{\bf\dot{r}} + \nabla_v I\cdot{\bf f}_{\rm geo}\equiv0$. The perturbation equations thus take the rather simple form
\be
\dot{I} = \nabla_vI \cdot \delta{\bf f}~.
\label{gpe}
\en
Given an explicit expression for  the perturbing force we can integrate these equations.

The osculating element method can be formulated in several different ways. There is freedom in the parameterization of the geodesic solution that is used as a basis for deriving the osculating element equations, and in the basis used to prescribe the force. It is also possible to treat the orbital phase constants either as constants of the motion that are evolved explicitly or as part of a total phase variable which satisfies new equations that  depend on the perturbation. We will describe two methods for treating the Kerr problem: (i) evolution of  $E, L_z, Q$ and the \emph{full} orbital phases with the force prescribed with respect to the Kinnersley orthonormal tetrad; (ii)  evolution of
 the orbital constants of motion $E, L_z, Q$ and the initial phases, with the force prescribed by its Boyer-Lindquist components.

In the Hamiltonian approach we start with an unperturbed Hamiltonian, $H_0$ and write the
equations of motion in terms of the constant canonical coordinates and momenta, $X^{\alpha},
P_{\alpha}$ (Hamilton-Jacobi approach), which are closely related, if not exactly the same,
as the six constants of motion introduced above, $I$ \cite{Goldstein}. If we can describe the perturbation as a small addition $\delta H$ to the unperturbed Hamiltonian, then we can describe the equations
of motion in the same generalized coordinates and momenta, which are no longer constants.
The derivatives of the perturbation  $\delta H$ give the equations for the
evolution of $X^{\alpha}, P_{\alpha}$. Quite often those equations are solved iteratively starting with an assumption
that the orbit is unperturbed in the right-hand side (in the $\delta H$). This is similar to the adiabatic solution to the osculating element equations which we will describe below. The Hamiltonian approach (if it can be formulated) would give equations equivalent to approach (ii) mentioned above.

We note that an obvious method of computing inspirals is to
numerically integrate the second-order forced geodesic equations
directly, taking the fundamental variables to be the Boyer-Lindquist
coordinates and their derivatives with respect to proper time.
The key advantage of the methods discussed in this paper over
second-order integrations is that they mesh much more naturally
with the adiabatic approximation and more generally with  two-time-scale
approximation techniques \cite{FH}.  For extreme-mass-ratio inspirals
driven by radiation reaction, the orbital evolution time scale is much
longer than the orbital time scale for most of the inspiral, until the
orbit becomes close to the innermost stable orbit.  The adiabatic
approximation to the motion gives the motion as an
expansion in the ratio of the time scales, and then there are various
postadiabatic corrections to this.  Although it is not possible yet to
compute numerically the full first-order self-force for generic orbits
in Kerr, it is possible to compute the averaged, dissipative piece of
this force, which is sufficient to compute leading-order adiabatic inspirals
\cite{FH}. The two-time-scale expansion also allows one to go beyond
the adiabatic evolution and compute the
small, rapidly oscillating perturbations to the evolution of the
orbital variables, as well as the slow secular changes to higher
order.

The two-time-scale method cannot be easily applied to the second-order,
forced geodesic equations, but it can be applied to the equations derived
in this paper, as we discuss in Secs.~\ref{NLosc} and \ref{Kerr}
below.  In particular, the osculating elements
method allows us to explicitly estimate the orbital average rate of change of
the orbital elements. This gives us
a physical insight into the effect of a perturbing force on the orbit
which is otherwise obscured in the integration of the second-order
equations of motion.  Estimation of these secular changes also allows us
to construct the adiabatic evolution of the orbit in the regime where
it is applicable.

\subsection{Numerical ``kludge'' waveform}
Another application for the results described in this paper is for the construction of numerical kludge waveforms. The numerical kludge waveform for EMRIs is a fast and accurate way to compute the long waveforms \cite{Babak:2006uv} that will be needed for EMRI data analysis. These are built in a not entirely consistent way, but the basic philosophy is to model the underlying trajectory of the inspiralling object as accurately as possible in order to obtain the best possible phase match between the true and approximate waveforms. The approximation is based on geodesic motion in the MBH's spacetime, combined with a flat spacetime  waveform generation expression. In the most recent version of the numerical kludge~\cite{Gair:2005ih}, the instantaneous geodesic orbit was updated by evolving the three constants of the motion $E, L_z, Q$ \cite{Glampedakis:2002cb} only. The evolution of the constants was obtained by combining post-Newtonian results with fits to numerical fluxes obtained by solving the Teukolsky equation \cite{Gair:2005ih}. However this method of evolving the  geodesics is not complete, as we described above, since we need to evolve the (initial) orbital phases  together with the orbital constants $E, L_z, Q$. In particular, the natural (and incorrect) way to evolve the phase constants, which is to fix them at some initial point, leads to significantly different evolutions in a time or frequency domain implementation of the kludge. The desire to resolve this apparent discrepancy between the two implementations was one of the initial motivations for the work described here. This article outlines the correct way to evolve geodesics under the self-force which could be used to further improve the numerical kludge waveforms in both time and frequency domain descriptions of them.

\subsection{Main results and the structure of the paper}
In this paper we will give a detailed description of the osculating elements approach applied to an arbitrary perturbing force acting on an object undergoing geodesic motion in the Kerr spacetime. As an introduction to the three dimensional relativistic problem of perturbed geodesic motion we will first consider a toy problem in Sec.~\ref{NLosc}. We look at the one-dimensional nonlinear oscillator acted on by an external force. The external force is chosen to have two components: a dissipative part and a conservative part (which just redefines the energy of the system). As we will see later this problem is a very good model for  the main problem of perturbed motion in the Kerr spacetime. We show how two implementations of the osculating elements approach work in this simplified model and compare the exact evolution with the adiabatic approximation. The
second of these two implementations [in which we evolve the energy and the initial time defined as $x(t_0) = 0$] allows us to treat the problem analytically in terms of Jacobi elliptic functions. This one-dimensional example allows the reader to understand the main approach which we then extend to the problem of forced geodesic motion in the Kerr spacetime in Sec.~\ref{Kerr}. We start that section with an introduction to our notation, before describing the osculating elements approach using the Kinnersley tetrad and ``Hughes'' variables (in terms of the orbital constants and the total phase variables).We then describe the forced geodesic equations in Boyer-Lindquist coordinates, evolving the orbital constants and the initial conditions, which is a direct extension to Kerr of the Schwarzschild results described in~\cite{pound08}. In both cases, we show how we can explicitly avoid the appearance of an apparent divergence in the osculating equations of motion at turning points.

In Sec.~\ref{gasdrag}, we illustrate our techniques with a problem in which the perturbing force is a ``gas-drag'' force proportional to the velocity of the inspiralling compact object. This is a toy model for an object inspiralling  in a gaseous environment around a MBH. We show that the different approaches give identical results, and once again compare the exact and adiabatic solutions to the problem. The influence of the drag force is to drive the inspiral of the object, but it also tends to increase the eccentricity of the orbit and
decrease the orbital inclination. Although we primarily use this problem for illustrative purposes, the increase in eccentricity is an interesting result that could have observational consequences. The increase in eccentricity is a purely relativistic effect, and is to be expected generically, as we discuss in more detail  in Appendix~\ref{A:SchDrag}, in the context of a drag force acting on an object in a Schwarzschild background.

We summarize and discuss our findings in the concluding section \ref{S:sum}. Some detailed mathematical calculations are included in additional appendixes.

\def\alt{
\mathrel{\raise.3ex\hbox{$<$}\mkern-14mu\lower0.6ex\hbox{$\sim$}}
}

\section{A simple model to illustrate methods used: the perturbed nonlinear oscillator}
\label{NLosc}

In this section we will study in detail the simple model of an
anharmonic oscillator subject to an external perturbing force, in
order to illustrate and explain in a simple context the methods that
we use for Kerr inspirals in subsequent sections of the paper.

We take the equation of motion for the position $x(t)$ of the oscillator to be
\be
{\ddot x} + x + \beta x^3 = \epsilon a_{\rm ext}(x,{\dot x})~.
\label{simplesystem}
\ee
Here the frequency of the oscillator is chosen to be unity for simplicity, $\beta>0$ is a parameter governing the size of the nonlinear term, $a_{\rm ext}$ is an externally applied perturbing acceleration, which could be a function of both the position and the velocity, and $\epsilon$ is a small parameter.  This simple system is similar in some respects to the system of a point particle in orbit about a Kerr black hole and subject to the gravitational self-force.  The dimensionless small parameter $\epsilon$ in the system (\ref{simplesystem}) plays the role of the mass ratio in the Kerr case, and the external acceleration $a_{\rm ext}$ is analogous to the self-force.

\subsection{Analysis using simple phase and energy coordinates on phase space}
\label{HughSol}

Consider initially the situation where the is no external
acceleration.  It is useful for some purposes to use a set of phase
space coordinates for the nonlinear oscillator which eliminate the
turning points.  We define coordinates $a$ and $\psi$, functions of
$x$ and $v \equiv {\dot x}$, by the equations
\bes
\bea
\frac{1}{2} a^2 +\frac{1}{4}  \beta a^4 &=& \frac{1}{2} {\dot x}^2 +\frac{1}{2}  x^2 + \frac{1}{4} \beta x^4~,\label{adef} \\
 \label{psidef}
x &=& a \cos \psi~,\\
{\rm sgn} ({\dot x}) &=& - {\rm sgn}( \sin \psi)~.
\label{signdef}
\eea
\ees
The expression on the right-hand side of Eq.\ (\ref{adef}) is just the
conserved energy of the system,
and $a$ is the conserved amplitude of the oscillation.
The variable $\psi$ increases monotonically (but not linearly) with
time.  The equations of motion in these variables are
\bes
\label{eom11}
\bea
{\dot a} &=& 0~, \\
\label{aeqn}
{\dot \psi} &=& \sqrt{1 + \beta a^2 (1 + \cos^2 \psi)/2}~.
\eea
\ees

Now consider turning on the external force.  Then the right-hand sides
of the equations of motion (\ref{eom11}) will acquire terms
proportional to $\epsilon$.  If we differentiate the definition
(\ref{psidef}) of $\psi$ with respect to $t$, insert the result into
the definition
(\ref{adef}) of
$a$, and solve for ${\dot \psi}$ using also Eq.\ (\ref{signdef}) we obtain
\be
{\dot \psi} =\frac{{\dot a}}{a} \cot \psi + \sqrt{1 + \beta a^2 (1 +
  \cos^2 \psi)/2}~,
\label{dotpsians}
\ee
which explicitly shows the extra forcing term.
However this term contains an apparent divergence at $\psi =0$.
The divergence is only apparent, since ${\dot a}$ will be constrained
to vanish when $\psi=0$, because the rate at which the force does
work will vanish when the velocity of the particle is zero.

To see this explicitly, we substitute the definition (\ref{psidef}) of
$\psi$ into the definition (\ref{adef}) of $a$, and solve for $\dot x$ to get
\be
\dot x = - a \sin\psi \sqrt{1 + \beta a^2 (1 + \cos^2 \psi)/2}~.
\label{dotxans}
\ee
Next, we differentiate both sides of Eq.\ (\ref{adef}) with respect to
$t$, and simplify the right-hand side using the equation of motion
(\ref{simplesystem}).  This gives
\be
(a + \beta a^3) {\dot a} = \epsilon {\dot x} a_{\rm ext}~.
\ee
Now using the result (\ref{dotxans}) for ${\dot x}$ and substituting
into Eq.\ (\ref{dotpsians}) gives the final results
\bes
\label{finalans}
\begin{align}
&{\dot \psi} =  \sqrt{1 + \beta a^2 (1 +  \cos^2 \psi)/2}
\left[ 1 - \epsilon \frac{\cos \psi a_{\rm ext}}{a
    (1 + \beta a^2)} \right]~,&\\
&{\dot a} = - \epsilon
 \sqrt{1 + \beta  a^2 (1 +  \cos^2 \psi)/2} \
\frac{\sin \psi \, a_{\rm ext}}{1 + \beta a^2}~,&
\end{align}
\ees
where $a_{\rm ext}(x,v)$ is evaluated at $x = a \cos \psi$, and $v =
v(a,\psi)$ given by the expression (\ref{dotxans}).

The final result (\ref{finalans}) now casts the system of differential
equations entirely in terms of the variables $a$ and $\psi$, and as
expected there are no divergences.  Note however that Eq.~(\ref{dotpsians}) would show a
divergence if one used an approximate, orbit-averaged version of
${\dot a}$ instead of the exact expression for ${\dot a}$.

In the analogous problem in Kerr, it is very straightforward to
compute the analog of the equation of motion (\ref{dotpsians}) which
contains the apparent divergence.  For numerical work, this form of
the equation would be problematic, since the right-hand side evaluates
to $0/0$ at turning points.  Our goal was to attempt to reformulate
the equations in Kerr analytically, to achieve a form analogous to Eq.\
(\ref{finalans}), where all the divergences have been removed.
Although it was not clear {\it a priori} that this would be possible
(because of the complexity of the Kerr-orbit dynamical system), we were
successful in finding an explicitly finite form of the equations of
motion in both sets of variables.

For the problem that we are really interested in, perturbed geodesics
in the Kerr
spacetime, it will be especially useful to consider the adiabatic
limit $\epsilon \to 0$ of small external perturbations.
So we consider adiabatic perturbations in the context of our example
problem.
The equations of motion (\ref{finalans}) for $\psi$ and $a$ can be
written in the general form
\bes
\label{generalform}
\bea
{\dot \psi} &=& \omega(\psi,a) + \epsilon g^{(1)}(\psi,a) +
O(\epsilon^2)~, \\
{\dot a} &=& \epsilon G^{(1)}(\psi,a) + O(\epsilon^2)~.
\eea
\ees
Here on the right-hand side, all the functions are periodic functions
of $\psi$ with period $2 \pi$.
In Appendix \ref{app:adiabatic} we derive the limiting form of the solutions in the
limit $\epsilon \to 0$; see also Ref. \cite{FH}.  The leading order or
adiabatic solutions are given by the following set of steps:
\begin{enumerate}
\item We define the averaging operation, for any function $f(\psi)$ of
  $\psi$, by
\be
\left< f \right>_a \equiv \frac{ \int_0^{2 \pi} d\psi
  \frac{f(\psi)}{\omega(\psi,a)}}{\int_0^{2 \pi} d\psi \frac{1}{\omega(\psi,a)}}~.
\ee
The subscript $a$ on the left hand side is a reminder that the
averaging operation depends on the value of $a$.

\item We define the averaged functions
\be
{\bar \omega}(a) \equiv \langle \omega(\psi,a) \rangle_a~,
\ee
and
\be
{\bar G}^{(1)}(a) \equiv \langle G^{(1)}(\psi,a) \rangle_a~.
\ee

\item We solve a pair of ordinary differential equations in the slow
  time parameter
\be
{\tilde t} = \epsilon t~,
\ee
for two auxiliary functions $\chi^{(0)}({\tilde t})$ and
$a^{(0)}({\tilde t})$.  This pair of ordinary differential equations is
\bes
\bea
\frac{d \chi^{(0)}}{d {\tilde t}} &=& {\bar \omega}(a^{(0)}({\tilde
  t}))~,\\
\frac{d a^{(0)}}{d {\tilde t}} &=& {\bar G}^{(1)}(a^{(0)}({\tilde
  t}))~.
\eea
\ees
Note that for this step, one does not need to specify a value of
$\epsilon$.

\item We can then write down the adiabatic solutions:
\bes
\label{adiabaticsolution}
\bea
a(t,\epsilon) &=& a^{(0)}(\epsilon t)~, \\
\psi(t,\epsilon) &=& \Xi\left[\frac{1}{\epsilon} \chi^{(0)}(\epsilon t)~,
a^{(0)}(\epsilon t) \right]~,
\eea
\ees
where the function $\Xi(\chi,a)$ is defined implicitly by the equation
\be
\frac{\chi}{2 \pi} = \frac{\int_0^{\Xi(\chi,a)}
  \frac{d\psi}{\omega(\psi,a)}}{\int_0^{2\pi}
  \frac{d\psi}{\omega(\psi,a)}}~.
  \label{bigxidef}
\ee
and satisfies $\Xi(\chi+2\pi,a) = \Xi(\chi,a) + 2\pi$.  (The inverse
of the mapping $\Xi$ essentially maps the given phase space
coordinates onto action-angle variables.)
\end{enumerate}

Note that there is an asymmetry in how the forcing terms $g^{(1)}$ and
$G^{(1)}$ in Eq.\ (\ref{generalform}) enter into the adiabatic
solution (\ref{adiabaticsolution}).
The function $G^{(1)}$, which drives the energy evolution, does enter,
but the function $g^{(1)}$, which drives the phase evolution, does not
enter at all.  It influences only the post-1-adiabatic solutions.

Note also that one cannot obtain the adiabatic
solutions by any simple modification of the original differential
equations.

\subsection{Analysis exploiting analytic solution to un-forced motion}
\label{EllSol}
It is also possible to find an analytic solution to the unperturbed anharmonic oscillator in terms of elliptic functions. Equation~(\ref{adef}) can be rearranged to give
\be \label{xdot}
\dot{x}^2 = \frac{\beta}{2} \left(x^2_+-x^2\right)\left(x^2-x^2_-\right) ~,
\ee
where we have defined the turning points
\be
x^2_{\pm} = \frac{1}{\beta}\left( -1 \pm \sqrt{1+2E\beta}\right)~.
\ee
in terms of the energy $E$, which is set to be twice the conserved quantity on the right-hand side of Eq.~(\ref{adef}), and is related to the amplitude of motion $a$ and the nonlinearity parameter $\beta$ by
\be
E = a^2+\frac{1}{2}\beta a^4 ~.
\ee
For $\beta > 0$, all of the solutions are bound and oscillate periodically in the interval $-x^2_+\leq x \leq x^2_+$. Without loss of generality we can set $x(t_0) = 0$, in which case
Eq.~(\ref{xdot}) can be rearranged and integrated to give
\begin{align} \label{solutions}
&\int_0^x \frac{{\rm d}y}{\sqrt{(y^2-x^2_-)(x^2_+-y^2)}} = \frac{1}{\sqrt{x^2_+-x^2_-}}&\nonumber \\
&\times\mathbf{F}\left[\sin^{-1}\left(\frac{x}{x_+} \sqrt{\frac{x^2_+-x^2_-}{x^2-x^2_-}}\right); \frac{x_+}{\sqrt{x^2_+-x^2_-}}\right] \nonumber \\
&= \pm\sqrt{\frac{\beta}{2}}(t-t_0)~,&
\end{align}
Here $\mathbf{F}(\phi;k)$ denotes the Jacobi elliptic integral of the first kind \cite{gnr}
\bea
\mathbf{F}(\phi;k)
&=& \int_0^{\sin\phi} \frac{dx}{\sqrt{(1-x^2)(1-k^2x^2)}} \nonumber \\
&=& \int_0^\phi\frac{d\alpha}{\sqrt{1 - k^2 \sin^2\alpha}}~.
\eea
In the following, we will denote all elliptic integrals by bold capital letters.
The inverse of this elliptic integral is given by the elliptic function ${\rm sn}(u;k)$
such that
\be
\mathbf{F}(\phi;k)=u \qquad\Leftrightarrow\qquad \phi(u;k) = \sin^{-1}\left[{\rm sn}(u;k)\right] \label{sndef}~,
\ee
For the solutions (\ref{solutions}), the parameter $k$ and argument $u$ are given by
\bea
k^2&=&\frac{x^2_+}{x^2_+-x^2_-} = \frac{\sqrt{1+2E\beta}-1}{2\sqrt{1+2E\beta}}~, \\
u&=&\left(1+2E\beta\right)^{1/4} (t-t_0)~,
\eea
The relation between $x(t)$ and the elliptic function is then
\be
\frac{x}{\sqrt{x^2-x^2_-}} = k {\rm sn}(u;k)~.
\ee
Solving this gives an expression for $x^2$ that is somewhat unsatisfying since using it requires manually
flipping the sign of $x$.  We can instead get a simpler expression if we introduce the
additional elliptic function
\be
{\rm dn}(u;k) = \sqrt{1-k^2 {\rm sn}^2(u;k)}~.
\ee
The result is
\be
x=k\sqrt{\frac{2(1-k^2)}{\beta(1-2k^2)}}{\rm sd}(u;k)~,
\label{xgeodef}
\ee
where ${\rm sd}(u;k) = {\rm sn}(u;k)/{\rm dn}(u;k)$.

We will now derive the osculating element equations for the variables $u$ and $k$. The physical variables $E$ and $t_0$ can be obtained from these simply via
\bea
E&=&\frac{2k^2(1-k^2)}{\beta(1-2k^2)^2}~,\\
\frac{{\rm d}E}{{\rm d}t} &=& \frac{4k}{\beta(1-2k^2)^3}\frac{{\rm d}k}{{\rm d}t}~,\\
t_0&=&t-\sqrt{1-2k^2} u~, \label{t0}\\
\frac{{\rm d}t_0}{{\rm d}t} &=& 1-\sqrt{1-2k^2} \frac{{\rm d}u}{{\rm d}t} + \frac{2ku}{\sqrt{1-2k^2}}
\frac{{\rm d}k}{{\rm d}t}~.
\eea
To derive the equations of motion in the osculating element form we need to differentiate ${\rm sd}(u;k)$ with respect to $u$ and $k$. This gives
\bea
\label{dsd}
\frac{\partial {\rm sd}}{\partial u}(u;k) &=& \frac{{\rm cn}(u;k)}{{\rm dn}^2(u;k)}~, \\
\frac{\partial {\rm sd}}{\partial k}(u;k) &=& \frac{u \;{\rm cn}(u;k)}{k\;{\rm dn}(u;k)}
- \frac{\mathbf{E}[\phi(u;k);k] {\rm cn}(u;k)}{k(1-k^2) {\rm dn}^2(u;k)} \nonumber \\
&+& \frac{k\;{\rm sn}(u;k)}{(1-k^2){\rm dn}(u;k)}~,
\eea
where we have introduced the elliptic function ${\rm cn}(u;k)$, which is defined by the analogue of Eq.~(\ref{sndef}) but with $\sin^{-1}$ replaced by $\cos^{-1}$, and where $\mathbf{E}(\phi;k)$ is the elliptic integral of the second kind \cite{gnr}:
\bea \label{ellipticE}
\mathbf{E}(\phi;k) &=& \int_0^{\sin\phi}dx~\sqrt{\frac{1-k^2 x^2}{1-x^2}} \nonumber \\
&=& \int_0^\phi d\alpha~ \sqrt{1-k^2\sin^2\alpha}~.
\eea
Since the parameter $k$ depends only on the energy, the evolution equation can be derived directly from the equation for the energy evolution, which follows by differentiation of Eq.~(\ref{adef}) and use of Eq.~(\ref{simplesystem}):
\be
\frac{{\rm d}E}{{\rm d}t} = 2\dot{x} \epsilon a_{\rm ext}~.
\ee
The evolution equation for $u$ follows from differentiating the orbit equation with respect to time and setting this equal to the velocity of the unperturbed orbit, which is given by Eq.~(\ref{xgeodef}):
\be
\frac{{\rm d}x}{{\rm d}t}_{\rm geo} = \sqrt{\frac{2(1-k^2)k^2}{\beta(1-2k^2)^2}} \; \frac{\partial {\rm sd}}{\partial u}~.
\ee
Putting these elements together we find the equations for the osculating evolution of the orbit
\bea
\label{kdot}
\frac{{\rm d}k}{{\rm d}t} &=& \epsilon a_{\rm ext} (x, \dot{x}) (1-2k^2)^2 \sqrt{\frac{\beta}{2}(1-k^2)}\frac{\partial{\rm sd}}{\partial u}~,  \\
\label{udot}
\frac{{\rm d}u}{{\rm d}t} &=& \frac{1}{\sqrt{1-2k^2}}
        -\epsilon a_{\rm ext}(x,\dot{x})(1 - 2k^2)\sqrt{\frac{\beta}{2}(1-k^2)} \nonumber \\
        &\times& \left[ \frac{1-2k^2+2k^4}{k(1-k^2)(1-2k^2)} {\rm sd}(u;k) +\frac{\partial{\rm sd}}{\partial k}\right] ~,
\eea
where the perturbing force is to be evaluated for the geodesic position and velocity,
\bea
\label{xofu}
x &=& \sqrt{\frac{2k^2(1-k^2)}{\beta(1-2k^2)}} {\rm sd}(u;k)~, \\
\label{dxofu}
\dot{x} &=& \sqrt{\frac{2k^2(1-k^2)}{\beta(1-2k^2)^2}} \frac{\partial{\rm sd}}{\partial u}~.
\eea

We can now derive the adiabatic approximation to the solution of Eqs.~(\ref{kdot}) and (\ref{udot})
following the steps described at the end of Sec.~\ref{HughSol}. Eqs.~(\ref{kdot}) and (\ref{udot}) have the same general form as $d\psi/dt$ and $da/dt$ in~Sec.~\ref{HughSol}.  That is, we can write them as
\bes
\label{generalformosc}
\bea
{\dot u} &=& \omega(u,k) + \epsilon g^{(1)}(u,k) +
O(\epsilon^2) \\
{\dot k} &=& \epsilon G^{(1)}(u,k) + O(\epsilon^2)~,
\eea
\ees
where we have now redefined the functions $\omega$, $g^{(1)}$, and $G^{(1)}$.  By comparing against the formula for $\dot u$, we find
\be
\omega(u,k) = \omega(k) = (1 - 2k^2)^{-1/2}~.
\ee
As a result, the averaging operation is greatly simplified:
\be
\left< f(u,k) \right>_k = \frac{1}{\mathcal{U}(k)} \int_0^{\mathcal{U}(k)} du~ f(u,k)~,
\ee
where $\mathcal{U}(k)$ is the period in $u$ for the general solution (\ref{xofu})
\be
\mathcal{U}(k) = 4\mathbf{F}(\pi/2,k) = 4\mathbf{K}(k)~.
\ee
Here $\mathbf{K}(k)$ is the complete elliptic integral of the first kind.
Note however that this period depends on
$k$, whereas, in the previous parametrization, the period in $\psi$ was simply $2\pi$.

As before, the two functions we wish to average are $\omega$ and $G^{(1)}$.  Since $\omega$ is independent of $u$,
\be
\bar \omega = \left<\omega \right>_k = \omega~.
\label{ombar}
\ee
To make further progress we must specify the  perturbing force. We take this to be
\be
a_{\text{ext}} = -\gamma \dot x + \delta x^2~.
\label{pertdef}
\ee
By substituting the Eqs.~(\ref{xofu}) and (\ref{dxofu}) for $x(u,k)$ and $\dot x(u,k)$ into this expression, we find, omitting the arguments for the elliptic functions, all of which depend on both $u$ and $k$,
\bea
G^{(1)}(u,k) = -\gamma k(1-k^2)(1-2k^2) \left(\frac{\partial {\rm sd}}{\partial u}\right)^2 \nonumber \\
+ \delta k^2 (1-k^2)^{3/2}(1-2k^2) \sqrt{\frac{2}{\beta}} {\rm sd}^2 \frac{\partial {\rm sd}}{\partial u}~.
\eea
The second term in this expression is symmetric about zero, and has period $\mathcal{U}$, so it vanishes
under the averaging operation.  The first term is also periodic, but it does not vanish under averaging since it is always positive.
Recalling the relation between $\partial{\rm sd}/\partial u$ and the other elliptic functions (\ref{dsd}),
\be
\bar G^{(1)} = \left< G^{(1)}\right>_k =  -\gamma k (1-k^2)(1-2k^2)\left< \frac{{\rm  cn}^2}{{\rm dn}^4}\right>_k~,
\ee
and exploiting the following identities,
\bea
{\rm sn}^2 + {\rm cn}^2 &=& 1~, \\
{\rm dn}^2 + k^2{\rm sn}^2 &=& 1~,
\eea
we can rewrite $\bar G^{(1)}$ as
\be
\bar G^{(1)} = \gamma(1-2k^2)\left[\frac{1-k^2}{k} \left[ (1-k^2)\left<{\rm dn}^{-4}\right>_k - \left<{\rm dn}^{-2}\right>_k \right]\right]~.
\ee
The averaging operations can be reduced to just one integral by using the identity \cite{gnr}
\bea
\int du~ {\rm dn}^m &=& \frac{1}{(m+1)(1-k^2)} \bigg[
k^2 {\rm dn}^{m+1} ~{\rm sn} ~{\rm cn} ~\nonumber\\
&+& (m+2)(2-k^2) \int du~ {\rm dn}^{m+2}\nonumber \\
&-& (m+3) \int du~{\rm dn}^{m+4} \bigg]~.\label{dnint}
\eea
The first term in square brackets vanishes on the ends of the interval $\{0,\mathcal{U}\}$, so
\bea
\int_0^\mathcal{U} du~ {\rm dn}^m &=& \frac{1}{(m+1)(1-k^2)} \nonumber \\
&\times& \left[ (m+2)(2-k^2) \int_0^\mathcal{U} du~ {\rm dn}^{m+2}\right. \nonumber \\
&-&\left. (m+3) \int_0^\mathcal{U} du~{\rm dn}^{m+4} \right]~.
\eea
which after using Eq.~(\ref{dnint}) gives
\be
\bar G^{(1)} = \gamma(1-2k^2)\left[\left( \frac{2}{3}\frac{2-k^2}{k} - \frac{1}{k} \right) \left<{\rm dn}^2\right>_k
- \frac{1-k^2}{3k}\right]~.
\ee
The average can be written as an elliptic integral using \cite{gnr}
\be
\int du~ {\rm dn}^2 = \mathbf{E}[\phi(u;k);k]~,
\ee
where $\mathbf{E}(u;k)$ is the elliptic integral of the second kind given in Eq.~(\ref{ellipticE}), and the amplitude function $\phi(u;k)$ is given by Eq.~(\ref{sndef}). This leaves us with the final result
\be
\bar G^{(1)} = \gamma(1-2k^2)\left[\left( \frac{2}{3}\frac{2-k^2}{k} - \frac{1}{k} \right)
\frac{\mathbf{E}(k)}{\mathbf{K}(k)}
- \frac{1-k^2}{3k}\right]~,
\label{Gbar}
\ee
where we have used $\mathcal{U} = 4\mathbf{K}(k)$, and we use $\mathbf{E}(k) = \mathbf{E}(\pi/2,k)$ to denote the complete elliptic integral of the second kind.

\subsection{Example force}
\begin{figure}[htb!]
\includegraphics[width=0.45\textwidth]{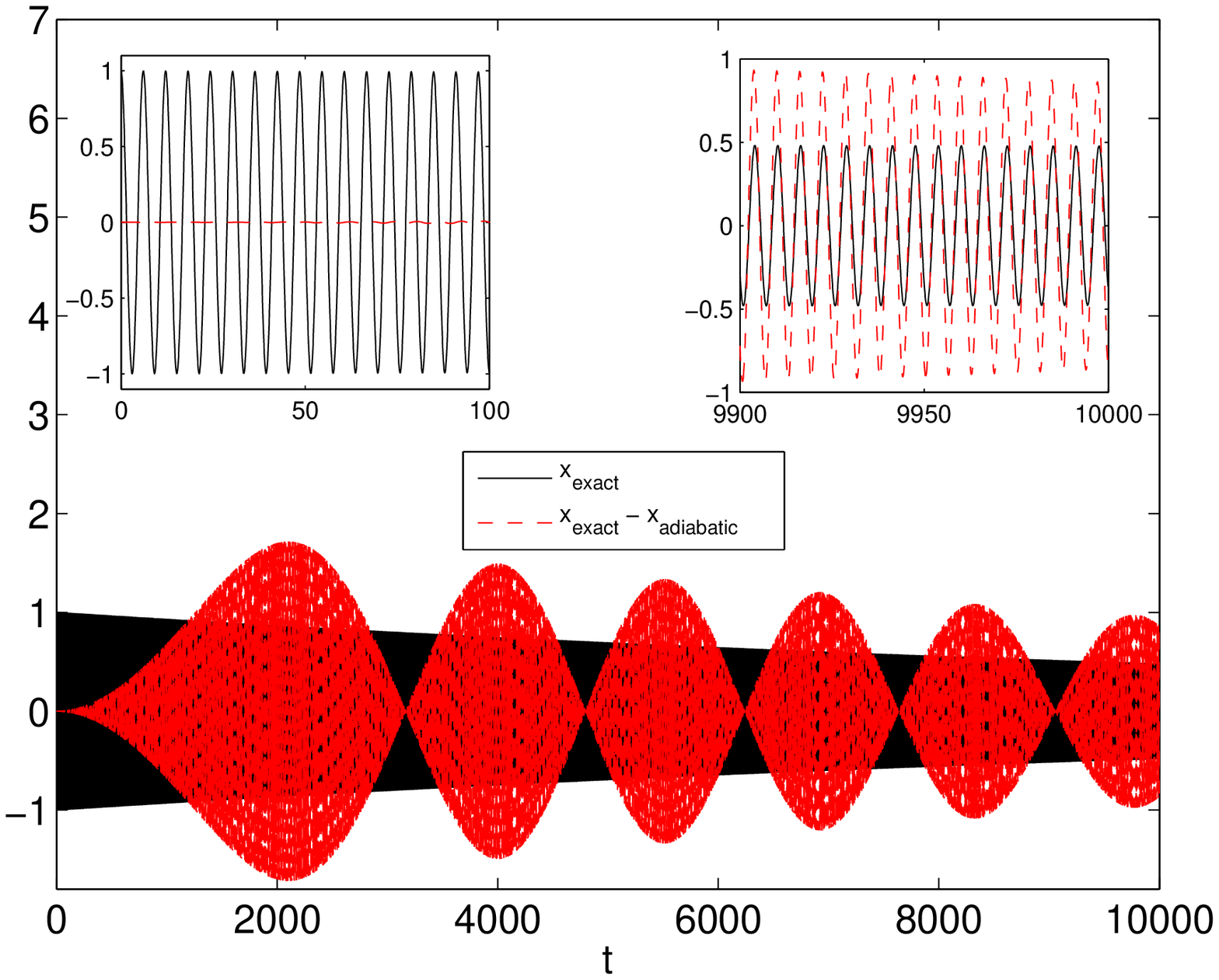}
\includegraphics[width=0.45\textwidth]{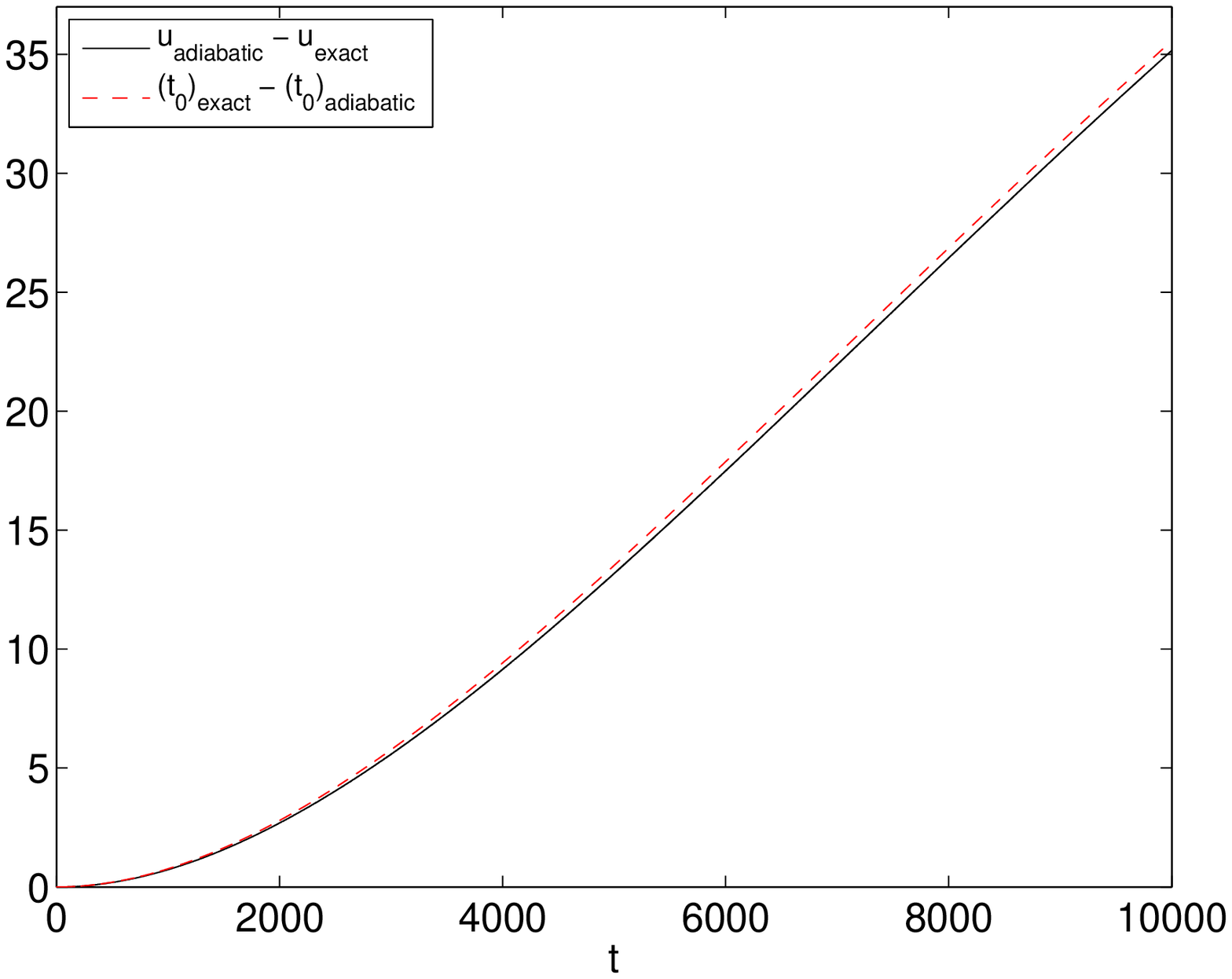}
\caption{Comparison between the exact and adiabatic approaches to evolving the orbit, both from
Sec.~\ref{EllSol}.  The top panel shows the exact solution $x(t)$ (solid lines), as well as the difference between the exact and adiabatic $x(t)$ (dashed lines).
The insets show close-up views of the first/last 100s of the same curves. The bottom panel shows the disagreement between the exact  and adiabatic predictions for the phase $u$ (solid line) and time-offset $t_0$ (dashed line).}
\label{OscComp}
\end{figure}

We will illustrate the techniques described above for an oscillator subject to the forcing term given in Eq.~(\ref{pertdef}), i.e.,
\be
a_{\rm ext} = - \gamma {\dot x} + \delta x^2
\ee
with $\epsilon = 10^{-3}$, $\beta = 0.1$, $\gamma = 0.15$, $\delta = 0.2$ and initial conditions $x(0) = 1.0$, ${\dot x}(0) = 0$.
The analytic solutions to the un-forced motion, as described in Secs.~\ref{EllSol} and \ref{HughSol}, were found to be essentially identical over the full integration time (as we would expect since these are both exact solutions to the forced motion).
In Fig.~\ref{OscComp}, we show the significant disagreements between
(i) using the analytic solution to the un-forced motion as described in Sec.~\ref{EllSol} (labeled ``exact''); and (ii) using the adiabatic approximation to the evolution, given by Eqs.~(\ref{generalformosc}), (\ref{ombar}) and~(\ref{Gbar}) (labeled ``adiabatic'').

There was no significant disagreement when it came to predicting the scale parameter $k^2$, but there was significant disagreement in what those formalisms predicted for the position $x(t)$, the phase $u$, and the time-offset $t_0$.
The top panel in Fig.~\ref{OscComp} shows that the adiabatic and exact positions go completely out of phase around $t=2000$, after which they then continue to pass in and out of phase with each other.  This is to be expected, since the adiabatic solution is only an approximation.  The bottom panel shows disagreements in both the phase $u$ and the time-offset $t_0$.  These grow to several cycles by the end of the integration, while the error in $k^2$ (not shown) remains small throughout the integration.  This is also to be expected.  Because of the terms we omit, we would expect the error in $k^2$ to scale like $\epsilon^2$, while the error in phase $u$, and correspondingly the error in $t_0$ from Eq.~(\ref{t0}), will scale like $\epsilon$.

Given that the exact solutions obtained via the analytic and phase solutions are the same, the choice of which parametrization to use must be made on the basis of practicality. The integration of the analytic form of the equations is more computationally expensive, as the elliptic functions must be evaluated at each integration step, so the phase form of the equations is probably preferable if we are interested only in the exact solution to $x(t)$. However, the averaged functions required for the adiabatic approximation to the solution are most easily derived from the analytic form of the equations, so this approach is better when we are interested in deriving an approximate solution to the equations.

\section{Osculating elements for orbits in the Kerr metric}
\label{Kerr}
\subsection{Summary of Notation}
\label{sec1}
In Boyer-Lindquist coordinates $(t,r,\theta,\phi)$, the Kerr metric is
\bea
ds^2 &=& - \left(1 - \frac{2 M r}{\Sigma}\right) dt^2
- \frac{4 a \sin^2 \theta M r}{\Sigma} dt d\phi \nonumber \\
&+& ( \varpi^4 - \Delta a^2 \sin^2 \theta)
\frac{\sin^2 \theta}{\Sigma} d\phi^2
 + \Sigma d\theta^2  \nonumber \\
&+&\frac{\Sigma}{\Delta} dr^2~.
\label{eq:kerrmetric}
\eea
Here
\bes
\label{eq:sigmaexpr}
\begin{eqnarray}
\Sigma & \equiv & r^2 + a^2 \cos^2 \theta~,  \\
\Delta & \equiv & r^2 + a^2 - 2 M r~, \\
\varpi & \equiv & \sqrt{r^2 + a^2}~,
\end{eqnarray}
\ees
and $M$, $a$ are the black hole mass and spin parameter.
Throughout the rest of this paper we use units in which $M=1$, for
simplicity.

We will make use of the Kinnersley null tetrad ${\vec l}$, ${\vec n}$,
${\vec m}$, ${\vec m}^*$, which is given by
\be
{\vec l} = \frac{\varpi^2}{\Delta} \partial_t + \partial_r +
\frac{a}{\Delta} \partial_\phi~,
\label{eq:vecldef}
\ee
\be
{\vec n} = \frac{\varpi^2}{2 \Sigma} \partial_t - \frac{\Delta}{2
  \Sigma} \partial_r + \frac{a}{2 \Sigma} \partial_\phi~,
\label{eq:vecndef}
\ee
and
\be
{\vec m} = \frac{1}{\sqrt{2}(r + i a \cos\theta)} \left( i a
\sin\theta \partial_t + \partial_\theta + \frac{i}{\sin\theta}
\partial_\phi \right)~.\label{eq:vecmdef}
\ee The corresponding one-forms are \be {\bf l} = - dt + a \sin^2
\theta d\phi + \frac{\Sigma}{\Delta} dr,\label{eq:dl} \ee
\be {\bf n} = - \frac{\Delta}{2 \Sigma} dt + \frac{a \Delta \sin^2
  \theta}{2 \Sigma} d\phi - \frac{1}{2} dr~,
\label{eq:dn}\ee
 and
 \be {\bf m} =  \frac{1}{\sqrt{2}(r + i a
\cos\theta)} \left( - i a \sin\theta dt + \Sigma d\theta + i
\varpi^2 \sin \theta d\phi \right)~. \label{eq:dm}\ee
The basis vectors obey the orthonormality relations ${\vec l} \cdot
{\vec n} = -1$ and ${\vec m} \cdot {\vec m}^* = 1$, while all other
inner products vanish.  The metric can be written in terms of the
basis one-forms as \be g_{\alpha\beta} = -2 l_{(\alpha} n_{\beta)} +
2 m_{(\alpha} m_{\beta)}^*~. \label{eq:gabformula} \ee

We define the conserved energy per unit rest mass $\mu$:
\be
E = - {\vec u} \cdot \frac{\partial}{\partial t}~,
\label{eq:Edef}
\ee
the conserved $z$-component of angular momentum divided by $\mu M$:
\be
L_z = {\vec u} \cdot \frac{\partial}{\partial \phi}~,
\label{eq:Lzdef}
\ee
and Carter constant divided by $\mu^2 M^2$:
\be
Q = u_\theta^2 - a^2 \cos^2 \theta E^2 +  \cot^2 \theta L_z^2 + a^2
\cos^2 \theta~.
\label{eq:Qdef}
\ee
(From now on we will for simplicity call these dimensionless quantities ``energy,''
  ``angular momentum,'' and ``Carter constant.'')
The geodesic equations can then be written in the form \cite{Drasco:2004tv}
\begin{align}
\left(\frac{dr}{d\lambda}\right)^2 &= \left[E(r^2+a^2)
- a L_z\right]^2- \Delta\left[r^2 + (L_z - a E)^2 +
Q\right]& \nonumber \\
&\equiv V_r(r)~,&
\label{eq:rdot}\\
\left(\frac{d\theta}{d\lambda}\right)^2 &= Q - \cot^2\theta L_z^2
-a^2\cos^2\theta(1 - E^2)& \nonumber \\
&\equiv V_\theta(\theta)~,&
\label{eq:thetadot}\\
\frac{d\phi}{d\lambda} &=
\csc^2\theta L_z + aE\left(\frac{r^2+a^2}{\Delta} - 1\right) -
\frac{a^2L_z}{\Delta}& \nonumber \\
&\equiv V_\phi(r,\theta)~,&
\label{eq:phidot}\\
\frac{dt}{d\lambda} &=
E\left[\frac{(r^2+a^2)^2}{\Delta} - a^2\sin^2\theta\right] +
aL_z\left(1 - \frac{r^2+a^2}{\Delta}\right)& \nonumber \\
&\equiv V_t(r,\theta)~.&
\label{eq:tdot}
\end{align}
Here $\lambda$ is the Mino time parameter \cite{Mino:2003yg}, related to
proper time $\tau$ by
\be
d\lambda = \frac{1}{\Sigma} d\tau~.
\label{eq:Minotime}
\ee
and we use these equations to define the potentials $V_r(r)$,
$V_\theta(\theta)$, $V_\phi(r,\theta)$ and $V_t(r,\theta)$.
Sometimes it will be convenient to use,
instead of the Carter constant $Q$, the quantity
\be
K = Q + (L_z - a E)^2~.
\label{Kdef}
\ee

For the rest of this section we specialize to bound geodesics, which are
periodic in $r$ and $\theta$.

\subsection{Change of Variables}
Eqs.~(\ref{eq:rdot}) -- (\ref{eq:tdot}) form a complete set of  equations that can be solved to
obtain the geodesic motion. However, it is difficult in practice to use the variables $r$ and
$\theta$ due to sign flips that occur in, for example,
$$
\frac{dr}{d\lambda} = \pm \sqrt{V_r(r)}~,
$$
at turning points.  Therefore we follow Drasco and Hughes in switching
to an alternative set of variables \cite{Drasco:2004tv,Babak:2006uv}.

\subsubsection{Angular motion}
We introduce the notation $z = \cos^2 \theta$, and note that the
effective potential can be written as
\be V_\theta(z) =
\frac{1}{1-z} \left[ Q(1-z) - z L_z^2 - \beta z (1-z)
\right]~,\label{eq:vthetaz1} \ee
where $\beta = a^2 (1-E^2)$.  We note that this $\beta$ is different from the variable appearing in the forced oscillator in Sec.~\ref{NLosc}. All subsequent references to $\beta$ will assume this new definition. We define $z_-$ and $z_+$ with $z_- <
z_+$ to be the two roots, so that \be V_\theta(z) = \frac{1}{1-z}
\beta(z_--z) (z_+-z)~, \label{eq:vthetaofz} \ee These roots $z_-$ and
$z_+$ are functions of $E$, $L_z$ and $Q$, and are positive with $0
< z_- < 1$ and $z_+>1$.  The motion takes place in the region $0 \le
z \le z_-$.

We replace the angular variable $\theta$, which oscillates, with
another angular variable $\psi_\theta$, which increases monotonically
with time.  The definition is given by
\be
\cos \theta = \sqrt{z_-} \cos \psi_\theta~.\label{eq:psithetadef}
\ee
Note that, for general forced motion $z_-$, will change with time,
along with $\theta$ and $\psi_\theta$.

\subsubsection{Radial motion}

We define $r_1$, $r_2$, $r_3$ and $r_4$ to be the roots of the radial
potential $V_r(r)$:
\be
V_r(r) = (1-E^2) (r_1 -r )(r - r_2) (r - r_3) (r - r_4)~.
\label{Vrdef}
\ee
Here the roots are ordered as $0 < r_4 < r_3 < r_2 < r_1$, and the
motion takes place in the region $r_2 < r < r_1$.  The roots are
functions of $E$, $L_z$ and $Q$.

We replace the radial variable $r$, which oscillates, with an angular
variable $\psi_r$, which increases monotonically with time.  The
definition is given by
\be
r = \frac{p} {1 + e \cos \psi_r}~,
\label{eq:psirdef}\ee
where the semilatus rectum $p$ and eccentricity $e$ are defined by
\be
r_1 = \frac{p}{1 - e}~, \ \ \ \ \ r_2 = \frac{p}{1 + e}~.\label{eq:rturningptdef}
\ee

\subsection{Forced motion using tetrad components of acceleration and using convenient phase and energy coordinates on phase space}
\label{tetform}

We now turn to the forced geodesic equation
\be
\frac{d^2 x^\alpha}{d\tau^2} + \Gamma^\alpha_{\beta\gamma} \frac{d
  x^\beta}{d\tau} \frac{d x^\gamma}{d\tau} = a^\alpha~,
\ee
where $a^\alpha$ is the external four acceleration.
In this subsection we derive our first formulation for integrating this
equation, which parametrizes the acceleration in terms of its
components on the Kinnersley null tetrad, and which parametrizes the
motion in terms of a convenient set of coordinates on phase space that
includes $E$, $L_z$ and $Q$.  The formulation is analogous to that
presented in Sec.\ \ref{HughSol} for the nonlinear oscillator model.
In particular, the phase variables used here are not conserved for
geodesic motion.  Our second formulation will be derived in the next
subsection.

Eqs.~(\ref{eq:rdot}) -- (\ref{eq:tdot}) are still valid for the
forced geodesic equation.  However, they must now be supplemented with
evolution equations for $E$, $L_z$ and $Q$ (or $K$).
We decompose the four acceleration on the Kinnersley tetrad as
\be
{\vec a} = -a_n {\vec l} - a_l {\vec n} + a_m^* {\vec m} + a_m {\vec
  m}^*~.\label{eq:adecomposition}
\ee
These four components are not all independent, since the acceleration
must be orthogonal to the four velocity.
We define
\be
R_a = \frac{1}{\sqrt{2}} (a_m + a_m^*)~, \ \ \ \ \
I_a = \frac{i}{\sqrt{2}} (a_m - a_m^*)~,
\ee
and we take the three
independent components to be $a_n$, $R_a$ and $I_a$.
In Sec.~\ref{tetBLconn} we will show how the tetrad components of the acceleration, $(a_n, a_l, a^*_m, a_m)$, relate to the acceleration components in Boyer-Lindquist coordinates, $(a_t, a_r, a_\theta, a_\phi)$.

We similarly decompose the four velocity in terms of the Kinnersley
tetrad as
\be
{\vec u} = -u_n {\vec l} - u_l {\vec n} + u_m^* {\vec m} + u_m {\vec
  m}^*~,
\ee
and we define
\be
R_u = \frac{1}{\sqrt{2}} (u_m + u_m^*)~, \ \ \ \ \
I_u = \frac{i}{\sqrt{2}} (u_m - u_m^*)~,
\ee
The components are given by the expressions
\begin{subequations}
\label{eq:velcomponents}
\begin{eqnarray}
u_l &=& u_r - \frac{F}{\Delta}~, \\
u_n &=& - \frac{F}{2 \Sigma} - \frac{\Delta}{2 \Sigma} u_r~, \\
R_u &=& \frac{r}{\Sigma} u_\theta + \frac{a {\cal H} \cos
  \theta}{\Sigma \sin \theta}~, \\
I_u &=& \frac{a \cos\theta}{\Sigma} u_\theta - \frac{r {\cal
    H}}{\Sigma \sin\theta}~.
\end{eqnarray}
\end{subequations}
where \be {\cal H} = L_z - a E \sin^2 \theta~,\label{eq:calHdef} \ee and \be F =
\varpi^2 E - a L_z~.\label{eq:Fdef} \ee The orthonormality condition ${\vec u} \cdot
{\vec a}=0$ allows us to solve for $a_l$: \be a_l = -
\frac{u_l}{u_n} a_n + \frac{1}{u_n}( R_a R_u + I_a
I_u)~.\label{eq:alexpr} \ee
We also define the following three
combinations of acceleration components:
\bes
\label{eq:calAdef}
\begin{eqnarray}
{\cal A}_I &=& r R_a + a I_a \cos\theta~, \\
{\cal A}_{II} &=& r I_a - a R_a \cos\theta~, \\
\label{eq:calAIII}
{\cal A}_{III} &=& R_u R_a + I_u I_a~.
\end{eqnarray}
\ees

We can now write down the evolution equations for the energy, angular
momentum and Carter constant.  These are (see Appendix \ref{appendix:derivation})
\be
\frac{dE}{d\lambda} = \frac{u_r a_n \Delta}{u_n} - \frac{
  \Delta {\cal A}_{III}}{2 u_n} - a \sin\theta {\cal
    A}_{II}~,
\label{eq:Eprime}
\ee
\be
\frac{d L_z}{d\lambda} = \frac{a \sin^2\theta u_r a_n \Delta}{
  u_n} - \frac{a \sin^2 \theta \Delta {\cal A}_{III}}{2  u_n} -
\varpi^2 \sin\theta {\cal A}_{II}~,
\label{eq:Lzprime}
\ee
and
\be
\frac{d K}{d\lambda} = 2 \Sigma^2 {\cal A}_{III}~.
\label{eq:Kprime}
\ee

\subsubsection{Equations of motion in terms of phase variables}
We next replace the equations of motion (\ref{eq:rdot}) and
(\ref{eq:thetadot}) for $r$
and $\theta$ with new equations of motion for $\psi_\theta$ and
$\psi_r$, which are derived in Appendix \ref{appendix:derivation}.
The new equation for $\psi_\theta$ is
\begin{align}
\frac{d\psi_\theta} {d \lambda} &= \sqrt{\beta(z_+ - z)} \left[ 1 +
  \frac{(1-z_-) \Sigma {\cal A}_I \cos\psi_\theta }{ \beta \sqrt{z_-}
    (z_+ - z_-) \sin\theta} \right]& \nonumber \\
&+ \frac{ \cos\psi_\theta \sin\psi_\theta {\cal H} a \Delta ( {\cal
    A}_{III} - 2 u_r a_n)}{2 (z_+ - z_-) \beta u_n} &\nonumber \\
&+ \frac{ \cos\psi_\theta \sin\psi_\theta {\cal G} {\cal
    A}_{II}}{\beta(z_+ - z_-)}~,& \label{eq:psithetaprime}
\end{align}
where $z = z_- \cos^2 \psi_\theta$ and
\be
{\cal G} = \frac{ \varpi^2 L_z}{\sin\theta} - a^3 (1-z_-) \sin\theta E~.
\ee
The new equation for $\psi_r$ is
\bea
\frac{d \psi_r}{d \lambda} &=& {\cal P}
+ \frac{{\cal C} {\cal A}_{III} \sin \psi_r}{2 (1 + e \cos\psi_r) u_n}
+ \frac{{\cal D} \Sigma {\cal A}_{III} {\cal P}}{2 (1 + e \cos \psi_r)^2 u_n} \nonumber \\
&-& \frac{a {\cal E} \sin\theta \sin\psi_r {\cal A}_{II} }{1 + e \cos \psi_r}
+ \frac{ {\cal P} a_n}{u_n (1 + e \cos \psi_r)^2} \nonumber\\
&\times& \left[ (1 -e)^2 (1 - \cos \psi_r) \frac{\Sigma_1 F_1}{\kappa_1} \right.\nonumber \\
&+&\left. (1+e)^2 (1 + \cos \psi_r) \frac{\Sigma_2 F_2}{\kappa_2} \right]~,\label{eq:psirprime}
\eea
where
\begin{widetext}
\begin{align}
{\cal P} &= p \sqrt{{\cal J}}/(1-e^2)~, & \\
  \label{eq:def_j}
  {\cal J} &=  (1-E^2)(1-e^{2})
                + 2\left(1-E^{2}-\frac{1-e^{2}}{p}\right)
                  (1+e\cos\psi_r)& \nonumber \\
               &+ \left\{(1-E^2)\frac{3+e^{2}}{1-e^{2}}-\frac{4}{p}+
                         \left[a^{2}(1-E^{2})+
                               L_{z}^{2}+
                               Q\right]\frac{1-e^{2}}{p^{2}}
                  \right\}
              (1+e\cos\psi_r)^{2}~,& \\
{\cal C} &= \frac{{\cal Q}_1 (1-e)}{\kappa_1} - \frac{{\cal Q}_2 (1+e)}{\kappa_2}~,& \\
{\cal D} &=  (1 - e)^2 (1 - \cos\psi_r) \frac{
    \Delta_1}{\kappa_1}
+ (1 + e)^2 (1 + \cos\psi_r) \frac{ \Delta_2}{\kappa_2}~,& \\
{\cal Q}_1 &= -2 a L_z r r_1-a^4 E
(r+r_1)+a^3 L_z (r+r_1)-a^2 E \left(r^3+r^2
  r_1+r_1^3+r r_1 (-2 + r_1)\right)& \nonumber \\
& -E
r r_1 \left(r r_1 (r+r_1)-2  \left(r^2+r
    r_1+r_1^2\right)\right)-a^2 \left(2 a^2 E -2
  E  r r_1+a L_z (-2 +r+r_1)\right)
\cos^2 \theta~,& \\
{\cal Q}_2 &= -2 a L_z r r_2-a^4 E
(r+r_2)+a^3 L_z (r+r_2)-a^2 E \left(r^3+r^2
  r_2+r_2^3+r r_2 (-2 + r_2)\right) &\nonumber \\
& -E
r r_2 \left(r r_2 (r+r_2)-2  \left(r^2+r
    r_2+r_2^2\right)\right)-a^2 \left(2 a^2 E -2
  E  r r_2+a L_z (-2 +r+r_2)\right)
\cos^2 \theta~,& \\
{\cal E} &= \frac{F_1 (1 - e) (r + r_1)}{\kappa_1} - \frac{F_2 (1 + e)
  ( r + r_2)}{\kappa_2}~.&
\end{align}
\end{widetext}
Here $\kappa = V'_r(r)$, and subscripts $1$ or $2$ mean that a
quantity is evaluated at $r = r_1$ or $r = r_2$ (except for ${\cal Q}_1$ and ${\cal Q}_2$).

\subsection{Forced motion using Boyer-Lindquist coordinate components
  of acceleration and phase variables that are conserved for
  geodesic motion}
\label{boyerform}

In this subsection we derive our second formulation for integrating
the forced geodesic equation, which is analogous to that presented in Sec.\ \ref{EllSol}
above for the nonlinear oscillator model.
This formulation parametrizes the acceleration in terms of its
Boyer-Lindquist coordinate components.  It parametrizes the motion in
terms of two phases $\psi_0$ and $\chi_0$ defined below, which are
conserved for geodesic motion, and three other parameters equivalent
to $E$, $L_z$ and $Q$, namely, the orbital eccentricity $e$, semilatus
rectum $p$ and angle of inclination $\iota$ (defined in Ref.\
\cite{Gair:2005ih}).
This formulation is a generalization of the
treatment of the Schwarzschild problem
described by Pound and Poisson in~\cite{pound08}.

In principle, we must evolve eight parameters, which are the four constants of motion and the four initial phase angles. However, one of these equations is eliminated by using the orthogonality condition $\dot{x^{\alpha}}a_{\alpha}=0$, where a dot denotes differentiation with respect to proper time, $\tau$, and $a^\alpha$ is the acceleration. This condition is discussed in~\cite{pound08} and comes from the definition of proper time.

In this section we will write the phase angles in the form $\psi_r = \psi-\psi_0$, $\psi_\theta=\chi-\chi_0$ and derive explicit equations for the time evolution of the initial-phase constants $\psi_0$ and $\chi_0$. The other parts of the phases, $\psi$ and $\chi$, are evolved using the standard geodesic expressions. While in practice we will need $\psi_r$ and $\psi_\theta$ to evolve the orbit, we decompose the equations this way to facilitate comparison to~\cite{pound08} and to make it easier to identify the conservative contributions from the perturbing force, which are essentially $\langle\dot{\psi_0}\rangle$, $\langle\dot{\chi_0}\rangle$.

\subsubsection{Contravariant formulation}
The Gaussian perturbation equations~(\ref{gpe}) carry over to the relativistic case and gives Eqs.~(27)-(32) in~\cite{pound08}. In the Kerr case, we have two additional equations as the $\theta$ motion is no longer trivial. In~\cite{pound08}, the equations were integrated with respect to the anomaly. In the Kerr case, as we have two anomalies, it will be more convenient to integrate the equations with respect to the coordinate time, $t$. The equations of motion that are independent of the force terms are
\bea
\frac{\partial r}{\partial p} p'+\frac{\partial r}{\partial e} e'+\frac{\partial r}{\partial \iota} \iota'+\frac{\partial r}{\partial \psi_0} \psi_0'+\frac{\partial r}{\partial \chi_0} \chi_0' &=& 0
~,\nonumber \\ && \label{kerrr}\\
\frac{\partial \theta}{\partial p} p'+\frac{\partial \theta}{\partial e} e'+\frac{\partial \theta}{\partial \iota} \iota'+\frac{\partial \theta}{\partial \psi_0} \psi_0'+\frac{\partial \theta}{\partial \chi_0} \chi_0' &=& 0
~,\nonumber \\ && \label{kerrtheta}\\
\frac{\partial \phi}{\partial p} p'+\frac{\partial \phi}{\partial e} e'+\frac{\partial \phi}{\partial \iota} \iota'+\frac{\partial \phi}{\partial \psi_0} \psi_0'+\frac{\partial \phi}{\partial \chi_0} \chi_0' +\Phi' &=& 0
~,\nonumber \\ &&\label{kerrphi}\\
\frac{\partial t}{\partial p} p'+\frac{\partial t}{\partial e} e'+\frac{\partial t}{\partial \iota} \iota'+\frac{\partial t}{\partial \psi_0} \psi_0'+\frac{\partial t}{\partial \chi_0} \chi_0' +T' &=& 0 ~.\nonumber \\ && \label{kerrt}
\ena
Here $\Phi$ and $T$ denote the phase offsets for the evolution of $\phi$ and $t$. We can ignore these equations if  we evolve $t$ and $\phi$ explicitly using the geodesic expressions evaluated along the instantaneous orbit, which amounts to evolving $t-T$ and $\phi-\Phi$  directly, as in the tetrad formulation. In the above, we use a dash to denote differentiation with respect to the parameter we use to define our orbit, which we take to be $t$. We will use a dot to denote differentiation with respect to the proper time $\tau$. The remaining four equations of motion are
\bea
\frac{\partial \dot{t}}{\partial p} p'+\frac{\partial \dot{t}}{\partial e} e'+\frac{\partial \dot{t}}{\partial \iota} \iota'+\frac{\partial \dot{t}}{\partial \psi_0} \psi_0'+\frac{\partial \dot{t}}{\partial \chi_0} \chi_0' &=& a^t \tau'
~,\nonumber \\ && \label{kerrtdot}\\
\frac{\partial \dot{r}}{\partial p} p'+\frac{\partial \dot{r}}{\partial e} e'+\frac{\partial \dot{r}}{\partial \iota} \iota'+\frac{\partial \dot{r}}{\partial \psi_0} \psi_0'+\frac{\partial \dot{r}}{\partial \chi_0} \chi_0' &=& a^r \tau'
~,\nonumber \\ && \label{kerrrdot}\\
\frac{\partial \dot{\theta}}{\partial p} p'+\frac{\partial \dot{\theta}}{\partial e} e'+\frac{\partial \dot{\theta}}{\partial \iota} \iota'+\frac{\partial \dot{\theta}}{\partial \psi_0} \psi_0'+\frac{\partial \dot{\theta}}{\partial \chi_0} \chi_0' &=& a^{\theta} \tau'
~,\nonumber \\ && \label{kerrthetadot}\\
\frac{\partial \dot{\phi}}{\partial p} p'+\frac{\partial \dot{\phi}}{\partial e} e'+\frac{\partial \dot{\phi}}{\partial \iota} \iota'+\frac{\partial \dot{\phi}}{\partial \psi_0} \psi_0'+\frac{\partial \dot{\phi}}{\partial \chi_0} \chi_0' &=& a^{\phi} \tau'
~.\nonumber \\ && \label{kerrphidot}
\ena
The terms $\partial\dot{r}/\partial p$ etc. denote differentiation of the geodesic equations given earlier with respect to the various orbital parameters. Following~\cite{pound08}, we can use the orthogonality condition to get rid of one of these equations, specifically Eq.~(\ref{kerrtdot}), and we will directly integrate $\phi$ and $t$ which means we do not need to consider Eqs.~(\ref{kerrphi})--(\ref{kerrt}).

We can rearrange Eqs.~(\ref{kerrr})--(\ref{kerrtheta}) to give
\bea
\psi_0' &=& -\frac{1}{\partial r/\partial\psi_0} \left( \frac{\partial r}{\partial p} p' + \frac{\partial r}{\partial e} e'  + \frac{\partial r}{\partial \iota} \iota' \right)~, \label{kerrpsi0}\\
\chi_0' &=& -\frac{1}{\partial \theta/\partial\chi_0} \left( \frac{\partial \theta}{\partial p} p' + \frac{\partial \theta}{\partial e} e'  + \frac{\partial \theta}{\partial \iota} \iota' \right)~, \label{kerrchi0}
\ena
where we have made use of the fact that the equation for $r$, (\ref{eq:rturningptdef}), is independent of $\chi_0$ and the equation for $\theta$, (\ref{eq:psithetadef}), is independent of $\psi_0$. The partial derivative $\partial r/\partial \iota$ also vanishes, but we include this term explicitly for simplicity of notation in the following. We generalize~\cite{pound08} by writing
\be
{\cal L}_a(x) = \frac{\partial \dot{x}}{\partial a} - \frac{\partial r/\partial a}{\partial r/\partial\psi_0} \frac{\partial\dot{x}}{\partial \psi_0} - \frac{\partial \theta/\partial a}{\partial \theta/\partial\chi_0} \frac{\partial\dot{x}}{\partial \chi_0}~.
\en
Substitution into Eqs.~(\ref{kerrrdot})--(\ref{kerrphidot}) then gives
\begin{widetext}
\bea
p' &=& \frac{\tau'}{D} \left( ({\cal L}_e(\theta){\cal L}_\iota(\phi)-{\cal L}_e(\phi){\cal L}_\iota(\theta))a^r + ({\cal L}_\iota(r){\cal L}_e(\phi)-{\cal L}_\iota(\phi){\cal L}_e(r))a^\theta + ({\cal L}_e(r){\cal L}_\iota(\theta)-{\cal L}_e(\theta){\cal L}_\iota(r))a^\phi \right)~,\\
e' &=& \frac{\tau'}{D} \left( ({\cal L}_\iota(\theta){\cal L}_p(\phi)-{\cal L}_\iota(\phi){\cal L}_p(\theta))a^r + ({\cal L}_p(r){\cal L}_\iota(\phi)-{\cal L}_p(\phi){\cal L}_\iota(r))a^\theta + ({\cal L}_\iota(r){\cal L}_p(\theta)-{\cal L}_\iota(\theta){\cal L}_p(r))a^\phi \right) \label{OscEcc}~,\\
\iota' &=& \frac{\tau'}{D} \left( ({\cal L}_p(\theta){\cal L}_e(\phi)-{\cal L}_p(\phi){\cal L}_e(\theta))a^r + ({\cal L}_e(r){\cal L}_p(\phi)-{\cal L}_e(\phi){\cal L}_p(r))a^\theta + ({\cal L}_p(r){\cal L}_e(\theta)-{\cal L}_p(\theta){\cal L}_e(r))a^\phi \right)~,\\
D &=& {\cal L}_p(r)({\cal L}_e(\theta){\cal L}_\iota(\phi)-{\cal L}_\iota(\theta){\cal L}_e(\phi))-{\cal L}_e(r)({\cal L}_p(\theta){\cal L}_\iota(\phi)-{\cal L}_\iota(\theta){\cal L}_p(\phi))+{\cal L}_\iota(r)({\cal L}_p(\theta){\cal L}_e(\phi)-{\cal L}_p(\phi){\cal L}_e(\theta))~.\nonumber\\&&
\ena
\end{widetext}
The correct evolution equations for the phase constants $\psi_0$ and $\chi_0$ may be found by substituting the preceding equations into (\ref{kerrpsi0})--(\ref{kerrchi0}). In the next section we will describe an alternative form of these equations which greatly simplifies the evolution of the constants of the motion. We include the above equations for completeness and to allow a direct comparison to the Schwarzschild results described in~\cite{pound08}.

\subsubsection{Covariant formulation}
The preceding section presented the equations in a contravariant formulation. We note that the equations for the evolution of the phase constants, (\ref{kerrpsi0})--(\ref{kerrchi0}), appear to be singular at turning points where $\partial r/\partial\psi_0=0$ or $\partial \theta/\partial\chi_0=0$. These are not real singularities, as the numerator also vanishes at the turning points, but it requires significant simplification to make this explicit. It is also possible to derive an alternative set of equations to (\ref{kerrtdot})--(\ref{kerrphidot}) from a covariant formulation of the equations. Pound and Poisson~\cite{pound08} chose the contravariant formulation in the Schwarzschild case, since they found it easier to eliminate the singularities at turning points in that formulation. However, there are advantages to using the covariant formulation, since two of the covariant velocity components are then equal to conserved quantities, $u_t=E$, $u_{\phi}=L_z$. The osculation conditions become
\begin{equation}
\frac{\partial x_G^\alpha}{\partial I^A} \dot{I}^A = 0~, \qquad \frac{\partial v^G_\alpha}{\partial I^A} \dot{I}^A = f_\alpha~,
\end{equation}
where
\be
v_\alpha^G = g_{\alpha\beta} \frac{{\rm d}z^\alpha_G}{{\rm d}\lambda}~,
\ee
in which $I^A$ denotes the orbital elements, including the phase constants. The first equation is the same as Eqs.~(\ref{kerrr})--(\ref{kerrt}) which reduce to (\ref{kerrpsi0}) and (\ref{kerrchi0}). The second equation is the equivalent of Eqs.~(\ref{kerrtdot})--(\ref{kerrphidot}), but in this case two of the equations simplify significantly, namely,
\begin{equation}
\dot{E}=f_t~, \qquad \dot{L_z}=f_{\phi}~. \label{covEv}
\end{equation}
In the Schwarzschild case, there is no equation for the $\theta$ motion and the radial equation follows from~(\ref{covEv}) through the constraint $\dot{z}^\alpha f_\alpha=0$. In the Kerr case, we do need to solve one of the radial or $\theta$ equations, or some combination of them. Alternatively, using the definition of the Carter constant in terms of the Killing tensor, we can derive the evolution equation for $Q$ straightforwardly.  The time evolution of the related constant $K$ defined in Eq.~(\ref{Kdef}), can be found from equation~(\ref{dKdt}) in appendix~\ref{appendix:derivation} as $\dot{K}=K^{\alpha\beta}u_\alpha a_\beta$. The Killing tensor $K^{\alpha\beta}$ can be written in terms of $l^\alpha$ and $n^\alpha$ as
\begin{equation}
K^{\alpha\beta} = 2\Sigma l^{(\alpha} n^{\beta)} + r^2 g^{\alpha\beta}~,
\label{Kofln}
\end{equation}
from which we obtain
\begin{equation}
\dot K = \dot E \frac{2}{\Delta} (\varpi^4 E - a \varpi^2 L_z)
       + \dot L_z \frac{2}{\Delta} (a^2 L_z - a\varpi^2 E)
       - 2\Delta u_r a_r~,
\end{equation}
where we have used $\dot E = -a_t$, and $\dot L_z = a_\phi$. An alternative expression for $K^{\alpha\beta}$ in terms of $m^\alpha$ and $m^{*\alpha}$ exists and is given in Appendix~\ref{appendix:derivation} as Eq.~(\ref{Kofmm}). If we had used this definition we would have found an equivalent expression for $\dot{K}$ that was a linear combination of $\dot E$, $\dot L_z$ and $a_\theta$.  The two expressions are equivalent, since the orthogonality relation between the perturbation force and four velocity always allows the elimination of one component of the force.

These three equations provide an alternative way to evolve the constants of the motion, $E$, $L_z$ and $Q$, but we must still evolve $\psi_0$ and $\chi_0$ using (\ref{kerrpsi0})--(\ref{kerrchi0}) and therefore we still need to deal with the turning points.

It is possible to derive an alternative form of these expressions that is manifestly finite at turning points by starting with the radial geodesic equation in the form
\begin{equation}
\Sigma^2 \dot{r}^2 = V_r(r,L_z,E,Q)~. \label{rdotgen}
\end{equation}
We need to show that the term
\begin{equation}
\frac{\partial r}{\partial E} \dot{E} + \frac{\partial r}{\partial L_z} \dot{L_z} + \frac{\partial r}{\partial Q}\dot{Q}~, \label{needcanc}
\end{equation}
that appears in an alternative version of Eq.~(\ref{kerrpsi0}), is proportional to $r'$. Differentiation of Eq.~(\ref{rdotgen}) with respect to $E$ yields
\begin{widetext}
\bea
2\Sigma^2\dot{r} \frac{\partial \dot{r}}{\partial E} + 2\Sigma\left(2r \frac{\partial r}{\partial E} -2a^2\cos\theta\sin\theta \frac{\partial \theta}{\partial E}\right)\dot{r}^2 =
\frac{\partial V_r}{\partial E} + \frac{\partial V_r}{\partial r} \frac{\partial r}{\partial E}~.
\eea
Similar equations may be obtained by differentiating with respect to $L_z$ and $Q$. Multiplying the $E$ equation by $\dot{E}$ etc. and adding the equations together, all terms on the left-hand side are proportional to $\dot{r}$, while on the right-hand side we get the expression (\ref{needcanc}) multiplied by $\partial V_r/\partial r$ plus the term
\begin{equation}
\frac{\partial V_r}{\partial E} \dot{E} + \frac{\partial V_r}{\partial L_z} \dot{L_z} + \frac{\partial V_r}{\partial Q} \dot{Q} = 2\dot{r} \Sigma^2\left( \ddot{r} - \frac{1}{2\Sigma^2}\frac{\partial V_r}{\partial r} + \frac{\dot{\Sigma}}{\Sigma} \dot{r}\right)~,
\end{equation}
where the second equality follows from differentiation of Eq.~(\ref{rdotgen}) with respect to time. The term in parentheses on the right-hand side is what we would obtain if we were on a geodesic, and therefore it necessarily equals $a^r$ in the evolving case. The final expression is
\bea
\dot{\psi_0} &=& 2\frac{\dot{\psi}_{\rm geo}}{\partial V_r/\partial r} \left[ \Sigma^2\left(\dot{E}\frac{\partial \dot{r}}{\partial E}+ \dot{L_z}\frac{\partial \dot{r}}{\partial L_z}+ \dot{Q}\frac{\partial \dot{r}}{\partial Q}\right) +2\Sigma r\dot{r}\left(\dot{E}\frac{\partial r}{\partial E} + \dot{L_z}\frac{\partial r}{\partial L_z}  + \dot{Q}\frac{\partial r}{\partial Q} \right)\right. \nonumber \\ &&\hspace{1in}\left. -2\Sigma a^2\cos\theta\sin\theta\dot{r}\left(\dot{E}\frac{\partial\theta}{\partial E} +\dot{L_z}\frac{\partial\theta}{\partial L_z}  +\dot{Q}\frac{\partial\theta}{\partial Q} \right)-\Sigma^2a^r\right]~,
\label{genphaseev}
\ena
\end{widetext}
in which $\dot{\psi}_{\rm geo}$ denotes the geodesic expression for $\rmd \psi/\rmd\tau$ which we use to evolve $\psi$. It is clear that this expression is indeed finite at radial turning points, provided that the radial self-force is finite. In the Schwarzschild case, it may be easily verified that Eq.~(\ref{genphaseev}) gives the same expression as Eq.~(\ref{kerrpsi0}) when they are explicitly simplified.

One important caveat is that although expression~(\ref{genphaseev}) is finite at radial turning points, it appears to diverge where $\partial V_r/\partial r = 0$, and this condition will be satisfied once between each consecutive turning point. This is not a real divergence either, which is clear from the fact that the original form of the equations did not show such a divergence. Therefore, if we were to substitute the various terms into the above expression we would find that the necessary cancellations would occur to eliminate these divergences. This simplification is a nontrivial calculation. However, an alternative approach that is easier to implement numerically is to use both Eqs.~(\ref{kerrpsi0}) and (\ref{genphaseev}) without any attempt to simplify the expressions. By switching from one expression to the other near turning points we can avoid numerical round-off problems. This is the implementation that we use in practice and from which the results presented in Sec.~\ref{gasdrag} were derived. We have verified in practice that both expressions do yield the same results at points where neither $V_r$ nor $\partial V_r/\partial r$ vanish.

\subsubsection{Action-angle formulation}
The method described above for evolving the equations of motion in the covariant formulation can be readily adapted to other problems and to other formulations of the Kerr geodesic solutions. In particular, an action-angle formulation of the Kerr solution exists~\cite{fujita09}, in which the equations take the form
\bea
X&=& A_X(E,L_z,Q) F_X(\psi_X-\psi_{X0};E,L_z,Q)~, \nonumber \\ && \\
\frac{{\rm d}\psi_X}{{\rm d}\lambda} &=& \Omega_X(E,L_z,Q)~,
\eea
where $X$ denotes $(t, r, \theta, \phi)$ and $\lambda$ is ``Mino time.'' The function $F_X$ is periodic for $r$ and $\theta$, with a period of $2\pi$\footnotemark, and for $t$ and $\phi$ it is the sum of a secular piece and an oscillatory term. The osculating element conditions give
\bea
&&\left(F_X \frac{\partial A_X}{\partial E} + A_X\frac{\partial F_X}{\partial E}\right) \frac{{\rm d}E}{{\rm d}\lambda} \nonumber \\
&+&
\left(F_X \frac{\partial A_X}{\partial L_z} + A_X\frac{ \partial F_X}{\partial L_z}\right) \frac{{\rm d}L_z}{{\rm d}\lambda} \nonumber \\
&+&\left(F_X \frac{\partial A_X}{\partial Q} + A_X\frac{\partial F_X}{\partial Q}\right) \frac{{\rm d}Q}{{\rm d}\lambda} = A_X F_X' \frac{{\rm d}\psi_{X0}}{{\rm d}\lambda}~,\nonumber \\ &&
\label{genosc}
\eea
where the dash denotes differentiation of $F_X$ with respect to the phase argument $\psi_X-\psi_{X0}$. As before, this expression appears to be singular at turning points, where $F'_X=0$. However, we can obtain an alternative expression by considering the potential
\be
\left(\frac{{\rm d}X}{{\rm d}\lambda}\right)^2 = V_X(X;E,L_z,Q)~.
\label{genpot}
\ee
Adding the derivative of this expression with respect to $E$ multiplied by ${\rm d}E/{{\rm d} \lambda}$ to the derivative with respect to $L_z$ multiplied by ${\rm d}L_z/{{\rm d} \lambda}$ and the derivative with respect to $Q$ multiplied by ${\rm d}Q/{{\rm d} \lambda}$ gives
\begin{widetext}
\bea
\frac{\partial V_X}{\partial X} \left(\left(F_X \frac{\partial A_X}{\partial E} + A_X\frac{\partial F_X}{\partial E}\right) \frac{{\rm d}E}{{\rm d}\lambda} +\left(F_X \frac{\partial A_X}{\partial L_z} + A_X\frac{ \partial F_X}{\partial L_z}\right) \frac{{\rm d}L_z}{{\rm d}\lambda}+\left(F_X \frac{\partial A_X}{\partial Q} + A_X\frac{\partial F_X}{\partial Q}\right) \frac{{\rm d}Q}{{\rm d}\lambda} \right) \nonumber \\
+ \frac{\partial V_X}{\partial E} \frac{{\rm d}E}{{\rm d}\lambda} +\frac{\partial V_X}{\partial L_z} \frac{{\rm d}L_z}{{\rm d}\lambda} +\frac{\partial V_X}{\partial Q} \frac{{\rm d}Q}{{\rm d}\lambda}  \nonumber \\
= 2\frac{{\rm d}X}{{\rm d}\lambda} \left(\frac{\partial}{\partial E} \left(\frac{{\rm d}X}{{\rm d}\lambda}\right) \frac{{\rm d}E}{{\rm d}\lambda} + \frac{\partial}{\partial L_z} \left(\frac{{\rm d}X}{{\rm d}\lambda}\right) \frac{{\rm d}L_z}{{\rm d}\lambda} + \frac{\partial}{\partial Q} \left(\frac{{\rm d}X}{{\rm d}\lambda}\right) \frac{{\rm d}Q}{{\rm d}\lambda} \right)~.
\eea

The derivative of Eq.~(\ref{genpot}) with respect to Mino time is
\be
2\frac{{\rm d}X}{{\rm d}\lambda} \frac{{\rm d}^2X}{{\rm d}\lambda^2} = \frac{\partial V_X}{\partial X} \frac{{\rm d}X}{{\rm d}\lambda} + \frac{\partial V_X}{\partial E} \frac{{\rm d}E}{{\rm d}\lambda} +\frac{\partial V_X}{\partial L_z} \frac{{\rm d}L_z}{{\rm d}\lambda} +\frac{\partial V_X}{\partial Q} \frac{{\rm d}Q}{{\rm d}\lambda}~,
\ee
which thus allows us to replace Eq.~(\ref{genosc}) with
\be
\frac{\partial V_X}{\partial X}  A_XF'_X \frac{{\rm d}\psi_{X0}}{{\rm d}\lambda} = \frac{{\rm d}X}{{\rm d}\lambda} \left( \left[  \frac{\partial V_X}{\partial X}  - 2 \frac{{\rm d}^2X}{{\rm d}\lambda^2} \right] +2 \left(\frac{\partial}{\partial E} \left(\frac{{\rm d}X}{{\rm d}\lambda}\right) \frac{{\rm d}E}{{\rm d}\lambda} + \frac{\partial}{\partial L_z} \left(\frac{{\rm d}X}{{\rm d}\lambda}\right) \frac{{\rm d}L_z}{{\rm d}\lambda} + \frac{\partial}{\partial Q} \left(\frac{{\rm d}X}{{\rm d}\lambda}\right) \frac{{\rm d}Q}{{\rm d}\lambda}  \right)\right)~.
\ee
\end{widetext}
The term in square brackets vanishes for geodesics and is therefore proportional to the $X$ component of the force when the orbit is perturbed. At turning points $F'_X$ and ${\rm d}X/{\rm d}\lambda$ are both zero and cancel, so we obtain a new form of the equation that is manifestly finite at turning points, albeit singular where ${\partial V_X}/{\partial X}=0$. As in the Boyer-Lindquist case, these two alternative formulations for the equations allow us to evolve the osculating element equations directly without worrying about singular behavior, just by switching between the two equivalent expressions in the vicinity of the turning points.\footnotetext{The choice of periodicity is in a sense arbitrary, and different periodicities could be obtained by rescaling the angular variable $\psi$. We specify a period of $2\pi$ for convenience.}

\subsection{Connection between Boyer-Lindquist and tetrad formulations}
\label{tetBLconn}
The tetrad formulation of the osculation equations, described in Sec.~\ref{tetform}, is written in terms of acceleration components, ${\cal A}_I$ etc., that are adapted to the Kinnersley tetrad, while the Boyer-Lindquist coordinate formulation, described in Sec.~\ref{boyerform}, is written in terms of the Boyer-Lindquist components of the acceleration. To identify the accelerations between the two approaches, we first write
down the tetrad components of the acceleration in terms of the Boyer-Lindquist components:
\begin{align}
a_n &= \frac{\varpi^2}{2\Sigma} a_t - \frac{\Delta}{2\Sigma} a_r + \frac{a}{2\Sigma} a_\phi ~,&\\
a_l &= \frac{\varpi^2}{\Delta} a_t + a_r + \frac{a}{\Delta} a_\phi ~,&\\
a_m &= \frac{1}{\sqrt{2}(r + i a \cos\theta)} \left(i a \sin \theta a_t + a_\theta + \frac{i}{\sin\theta} a_\phi \right) ~,&\\
a_m^* &= \frac{1}{\sqrt{2}(r - i a \cos\theta)} \left(-i a \sin \theta a_t + a_\theta - \frac{i}{\sin\theta} a_\phi \right)~.&
\end{align}
The acceleration functions $R_a=(a_m+a^*_m)/\sqrt{2}$ and $I_a=i(a_m-a^*_m)/\sqrt{2}$ introduced in Sec.~\ref{tetform} have components
\bea
R_a &=& \frac{a^2 \sin\theta \cos\theta}{\Sigma} a_t + \frac{r}{\Sigma} a_\theta + \frac{a\cot\theta}{\Sigma} a_\phi~, \\
I_a &=& -\frac{a r \sin\theta}{\Sigma}  a_t - \frac{a\cos\theta}{\Sigma} a_\theta - \frac{r}{\sin\theta\Sigma}a_\phi~,
\eea
from which we obtain the tetrad acceleration components in terms of the Boyer-Lindquist components of the acceleration
\begin{align}
{\cal A}_{I} &= \frac{r^2 - a^2\cos^2\theta}{\Sigma} a_\theta~,&\\
{\cal A}_{II} &= -a\sin\theta a_t - \frac{2ra\cos\theta}{\Sigma}a_\theta
- \frac{1}{\sin\theta} a_\phi~,& \\
{\cal A}_{III} &= -\alpha a \sin\theta a_t &\nonumber \\
&+ \frac{u_\theta(r^2 - a^2\cos^2\theta)/\Sigma - 2\alpha r \cos\theta}{\Sigma} a_\theta
 -\frac{\alpha}{\sin\theta} a_\phi~,&
\end{align}
in which
\be
\alpha = \frac{aE \sin^2\theta - L_z}{\Sigma\sin\theta}~.
\ee
In Sec.\ \ref{gasdrag} below we will consider a toy problem as an illustration of the two methods. The force will be specified in Boyer-Lindquist coordinates, and the preceding expressions can be used to obtain the corresponding tetrad components.

\subsection{Features and drawbacks of the two formulations}

In this final subsection we discuss some of the advantages and
disadvantages of our two formulations.

First, as discussed in the Introduction, earlier work on methods of computing
radiation reaction driven
inspirals focused on the adiabatic limit
\cite{hughescirc1,hughescirc2,Babak:2006uv,Glampedakis:2002cb,Gair:2005ih}.  In this limit, it is
sufficient to use orbit-averaged forces, or, equivalently,
orbit-averaged proper time derivatives of the first integrals, ${\dot
  E}$,
${\dot L}_z$ and ${\dot Q}$.  These quantities can be computed as
functions of $E$, $L_z$ and $Q$, both in post-Newtonian expansions and
exactly using numerical black hole perturbation theory.
In this paper our focus is on developing methods that allow going
beyond the adiabatic limit.  For this purpose, orbit-averaged
quantities are insufficient; one must use a prescription for the
perturbing force that depends on the two nontrivial orbital phases.
One could, in principle, continue to use the quantities ${\dot E}$,
${\dot L}_z$ and ${\dot Q}$ to parametrize the force, if these
quantities are taken to be functions of $E$, $L_z$ and $Q$ and of two
additional phases.  This would be the most natural way to generalize
the analyses of Refs.\ \cite{hughescirc1,hughescirc2,Babak:2006uv,Glampedakis:2002cb,Gair:2005ih}.

However, such a parametrization turns out to have a significant
disadvantage compared to the parametrizations used in
this paper, when one is attempting to compute approximate inspirals.
Specifically, there are constraints that the fluxes must
satisfy at radial and polar turning points, in order to ensure that
the four acceleration be finite.  Approximate versions of the fluxes
may violate the constraints and lead to cusps in the motion at the
turning points. (This will be true, in particular, for orbit-averaged
fluxes.)  The existence of these constraints can be seen from the
expression for the square of the four acceleration in terms of ${\dot
  E}$, ${\dot L}_z$ and ${\dot K}$, which is
\bea
{\vec a}^2 &=& \frac{1}{\Sigma \Delta u_r^2} \left( \frac{1}{2} {\dot K}
  - \frac{F {\bar F}}{\Delta} \right)^2
+ \frac{1}{\Sigma u_\theta^2} \left(\frac{1}{2} {\dot K} - {\cal G}
  {\bar {\cal G}} \right)^2 \nonumber \\
&& - \frac{ {\bar F}^2}{\Sigma \Delta} + \frac{ {\bar {\cal G}}^2}{\Sigma}.
\eea
Here $F = \varpi^2 E - a L_z$, ${\bar F} = \varpi^2 {\dot E} - a {\dot
  L}_z$, ${\cal G} = a \sin\theta E - \csc \theta L_z$, and
 ${\bar {\cal G}} = a \sin\theta {\dot E} - \csc \theta {\dot L}_z$.
It can be seen that at radial turning points where $u_r = 0$, the
fluxes must satisfy
the constraint ${\dot K} = 2 F {\bar F}/\Delta$, while at polar
turning points the constraint is ${\dot K} = 2 {\cal G} {\bar {\cal
    G}}$. \footnote{A similar phenomenon occurred in the nonlinear
  oscillator model of Sec.\ \ref{NLosc}, where the time derivative of
  the energy was constrained to vanish at turning points.}

By contrast, in the tetrad formulation used here, the magnitude of the
acceleration is automatically finite.
The independent components of the
four acceleration are taken to be three of the four components on the
Kinnersley null tetrad, namely $a_n$, $a_m$ and $a_m^*$, with the
fourth component being determined by the orthogonality of the
four acceleration and the four velocity.  In terms of these three
components, the square of the four acceleration is
\be
{\vec a}^2 = \frac{4 u_l a_n}{1 + 2 |u_m|^2} \left[ u_l a_n - u_m^*
  a_m - u_m a_m^* \right] + 2 |a_m|^2,
\ee
which is clearly always finite.\footnote{A related issue is that the time derivative of the orbital
eccentricity $e$ can diverge $ \propto 1/e$ as $e\to 0$, for forces
parameterized in terms of ${\dot E}$, ${\dot L}_z$ and ${\dot Q}$,
unless the fluxes obey certain constraints at $e=0$.  This issue is
discussed in detail in Ref.\ \cite{Gair:2005ih}.  Again, this
divergence is automatically excluded if one parameterizes the force in
terms of its tetrad components:  The eccentricity can be written as a
smooth function $e = e(x^\alpha,p^\beta)$ on phase space.  Taking a
proper time derivative gives $de/d\tau = m a^\alpha \partial
  e/\partial p^\alpha$, which is finite for finite accelerations.}

Similarly, in our Boyer-Lindquist formulation, the acceleration is
again always finite, except in some special cases in the ergosphere.
The independent components of the acceleration are taken to be the
spatial, contravariant components $a^i = (a^r, a^\theta,a^\phi)$, with
$a^t$ being determined by orthonormality.  The square of the
four acceleration is then
\be
{\vec a}^2 = \left( g_{ij} - 2 \frac{g_{ti} u_j}{u_t} + g_{tt}
  \frac{u_i u_j}{u_t^2} \right) a^i a^j,
\ee
which is always finite except in the ergosphere where it is possible
for $u_t$ to vanish.

We now turn to a discussion of a second issue, which is
a significant advantage of the Boyer-Lindquist formulation
over the tetrad
formulation.  This advantage is its simple behavior
under the discrete symmetries of the Kerr spacetime.  Specifically,
note that
any four acceleration ${\vec a} = {\vec a}(x^\alpha, u^\beta)$ can be
uniquely decomposed as the sum of a dissipative piece and a
conservative piece.  For the dissipative piece, the components $a^r$
and $a^\theta$ are odd under $u^r \to - u^r$, $u^\theta \to
-u^\theta$, while the components $a^t$ and $a^\phi$ are even.  For the
conservative piece, the components $a^r$ and $a^\theta$ are even,
while the components $a^t$ and $a^\phi$ are odd \cite{FH}.
It follows that, in the Boyer-Lindquist formulation, wherein one
specifies the
components $a^r$, $a^\theta$ and $a^\phi$ of the four acceleration, it
is straightforward to independently specify the dissipative and
conservative pieces.

By contrast, in the tetrad formulation presented here, the independent
variables are taken to be $a_n$, $a_m$ and $a_m^*$, and the
decomposition into conservative and dissipative pieces in terms of
these variables is somewhat involved.
In particular, if one is attempting to find useful approximations to
the conservative self-force, for example, by naively using
conservative post-Newtonian approximations to the quantities $a_n$,
$a_m$ and
$a_m^*$, the errors in the approximation will generically lead to a
self-force with both conservative and dissipative pieces.  This can be
a problem since in the adiabatic limit the effect of the dissipative
self-force on the motion is boosted relative to the conservative
self-force.

There are alternative parametrizations of the self-force that combine
the advantages of our two formulations, for example,
\be
a^\alpha = a^{{\hat r}} e_{{\hat r}}^\alpha + a^{{\hat \theta}} e_{{\hat \theta}}^\alpha + a_\perp \epsilon^{\alpha}_{\ \beta\gamma\delta} u^\beta e^\gamma_{{\hat r}} e^\delta_{{\hat\theta}} + (a^{{\hat r}} u_{{\hat r}} + a^{{\hat \theta}} u_{{\hat \theta}}) u^\alpha,
\ee
where ${\vec e}_{\hat r}$ and ${\vec e}_{\hat \theta}$ are unit
vectors in the directions of $\partial_r$ and $\partial_\theta$.  Here
the dissipative and conservative pieces of the quantities $a^{\hat
  r}$, ${\hat a}^{\hat \theta}$ and $a_\perp$
have simple transformation properties under discrete symmetries, and
moreover the magnitude of the four acceleration is
\be
{\vec a}^2 = (a^{\hat r})^2 \left[ 1 + u_{\hat r}^2 \right]
+ (a^{\hat \theta})^2 \left[ 1 + u_{\hat \theta}^2 \right]
+ a_\perp^2 \left[ 1 + u_{\hat r}^2 + u_{\hat \theta}^2 \right],
\ee
which is always finite.  Useful approximation schemes can be obtained
by (i) formulating approximations in terms of the
three variables $a^{\hat r}$, $a^{\hat \theta}$ and $a_\perp$; (ii)
using the exact, Kerr relations to compute $a_n$,
$a_m$ and $a_m^*$ in terms of $a^{\hat r}$, $a^{\hat \theta}$ and
$a_\perp$; and (iii)
using the resulting expressions in the tetrad formulation equations of
motion (\ref{eq:phidot}), (\ref{eq:tdot}), (\ref{eq:Eprime}) --
(\ref{eq:Kprime}), (\ref{eq:psithetaprime}) and (\ref{eq:psirprime}).
See Ref.\ \cite{2010arXiv1009.4923F} for an application of this approach.

\section{Example of perturbed Kerr Geodesics: ``gas-drag''}
\label{gasdrag}

As an example problem, we will suppose that the small mass experiences a drag force proportional to velocity, which could represent the behavior of an EMRI occurring in the presence of gas.  Here we derive the four acceleration for such a force.

In a given frame of reference, the relativistic analog of this simple drag force will have a term proportional to the spatial part of the velocity, plus a term proportional to the frame velocity constructed so that the force remains orthogonal to the total velocity.
Let $\vec{u}_\text{ZAMO}$ be the velocity of zero-angular-momentum observers (ZAMOs), and let $\vec{u}$ be the velocity of
the small mass.  In the frame of a ZAMO, the spatial part of the velocity of the small mass is
\be
\vec{u}_\perp = \vec{u} + \Gamma \vec{u}_\text{ZAMO}~,
\ee
where $\Gamma = \vec{u} \cdot \vec{u}_\text{ZAMO}$.  The drag force then has the form
\be\label{drag force}
\vec{a} = -\gamma \vec{u}_\perp + \kappa \vec{u}_\text{ZAMO} -\gamma \vec{u} + (\kappa -\gamma \Gamma)\vec{u}_\text{ZAMO}~,
\ee
where $\gamma$ is the linear drag coefficient.  Enforcing the condition
$\vec{a} \cdot \vec{u} = 0$ then determines the value of $\kappa$
\be
\kappa = \frac{\gamma(\Gamma^2-1)}{\Gamma}~.
\ee
Inserting this into the formula for the acceleration due to drag (\ref{drag force}) gives
\be
\vec{a} = -\gamma \left(\vec{u} + \frac{\vec{u}_\text{ZAMO}}{\vec{u}\cdot\vec{u}_\text{ZAMO}}\right)~.
\ee
Writing this explicitly in terms of Boyer-Lindquist coordinates, we have
\be
a^\alpha = -\gamma \left(u^\alpha + \frac{u_Z^\alpha}{{u_Z}_t u^t}\right)~,
\ee
where $u^\alpha$ denotes the four velocity of the inspiraling object and
\begin{align}
u_Z^t &=\sqrt{\frac{(r^2+a^2)^2-\Delta a^2\sin^2\theta}{\Sigma\Delta}}~,& \\
u_Z^\phi &= \frac{2ar}{\sqrt{\Sigma\Delta((r^2+a^2)^2-\Delta a^2 \sin^2\theta)}}~,&\\
u_Z^r &= u_Z^\theta = 0~,& \\
{u_Z}_t &= -\left(1-\frac{2r}{\Sigma}\right)u_Z^t - \frac{2a\sin^2\theta r}{\Sigma} u_Z^\phi ~,&
\end{align}
in which $\Delta = r^2-2Mr+a^2$, $\Sigma=r^2+a^2\cos^2\theta$ as before.

As a test case, we constructed an inspiral into a central black hole with spin $a=0.9$ under the influence of this gas-drag force with $\gamma=10^{-5}$. We took the initial orbital parameters to be $p/M=7$, $e=0.5$, $\iota=\pi/6$, $\phi=t=0$, $\psi_r=1$ and $\psi_\theta = 2$. The inspiral trajectory was constructed using both the Boyer-Lindquist and the tetrad formulations. The evolutions were found to be identical, as we would hope, and this gives us confidence that our results are correct. In the following discussion, we will not distinguish between the results obtained using the different formulations as they differed only at the level of numerical noise.

\begin{figure*}
\includegraphics[width=0.45\textwidth]{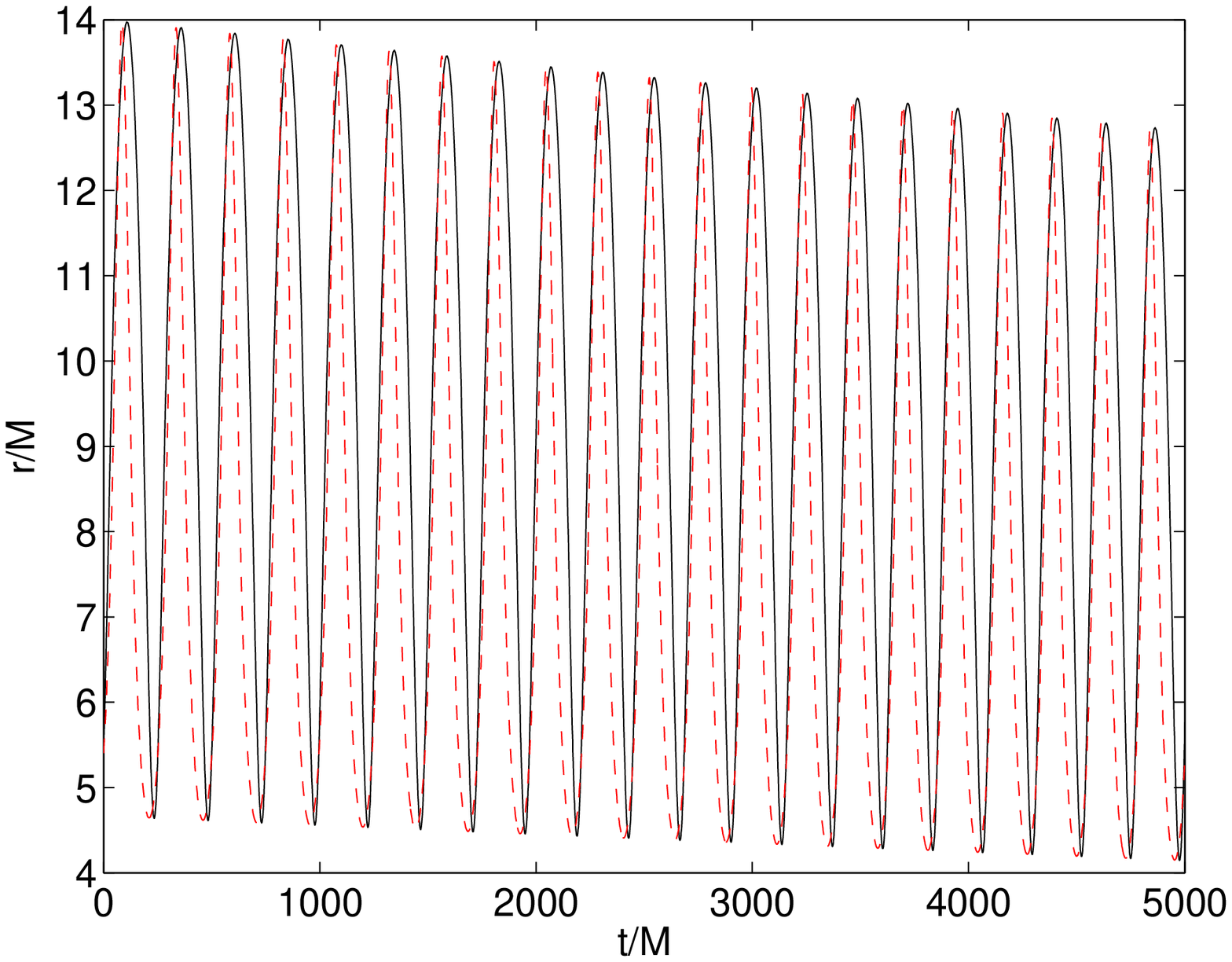}
\includegraphics[width=0.45\textwidth]{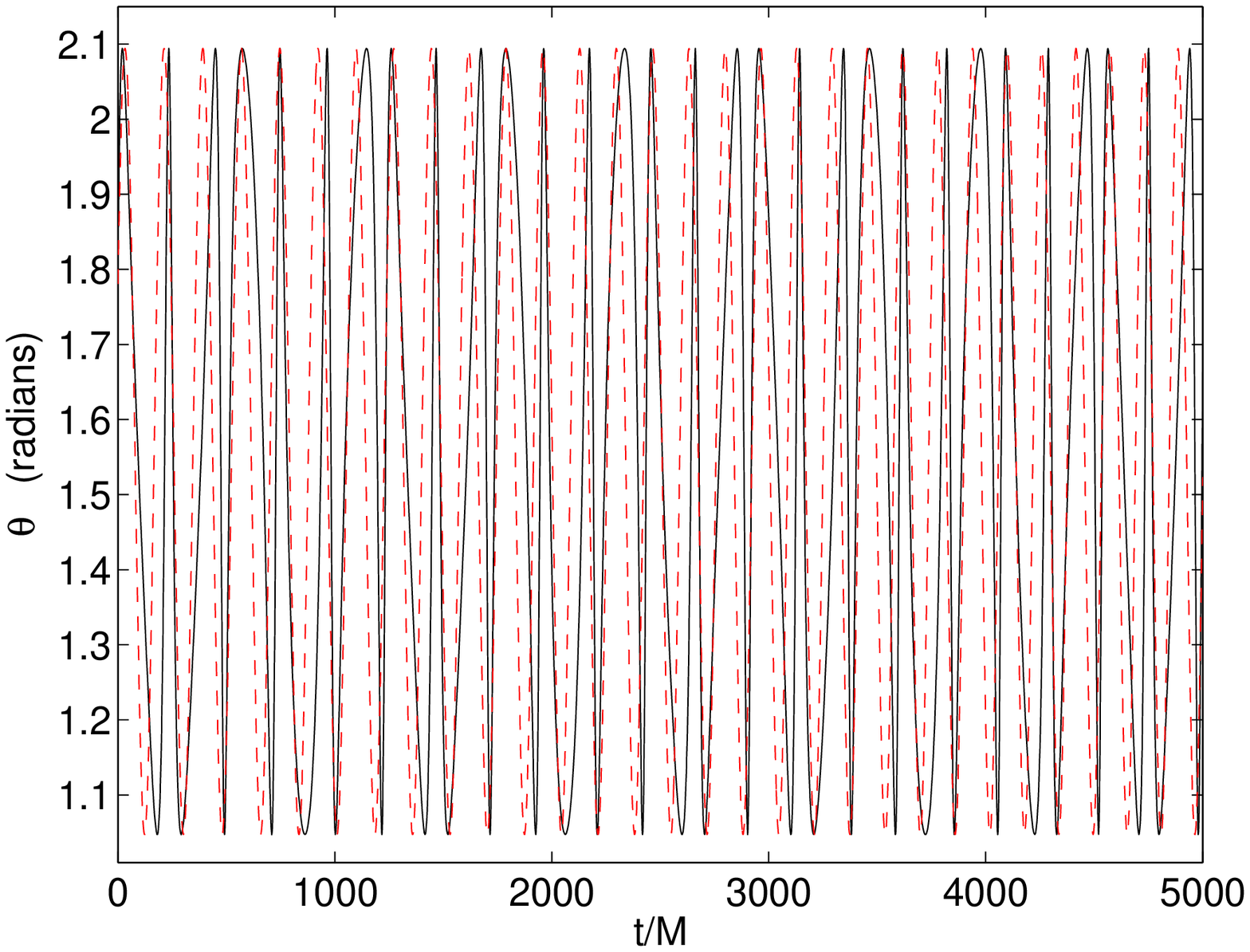}
\includegraphics[width=0.45\textwidth]{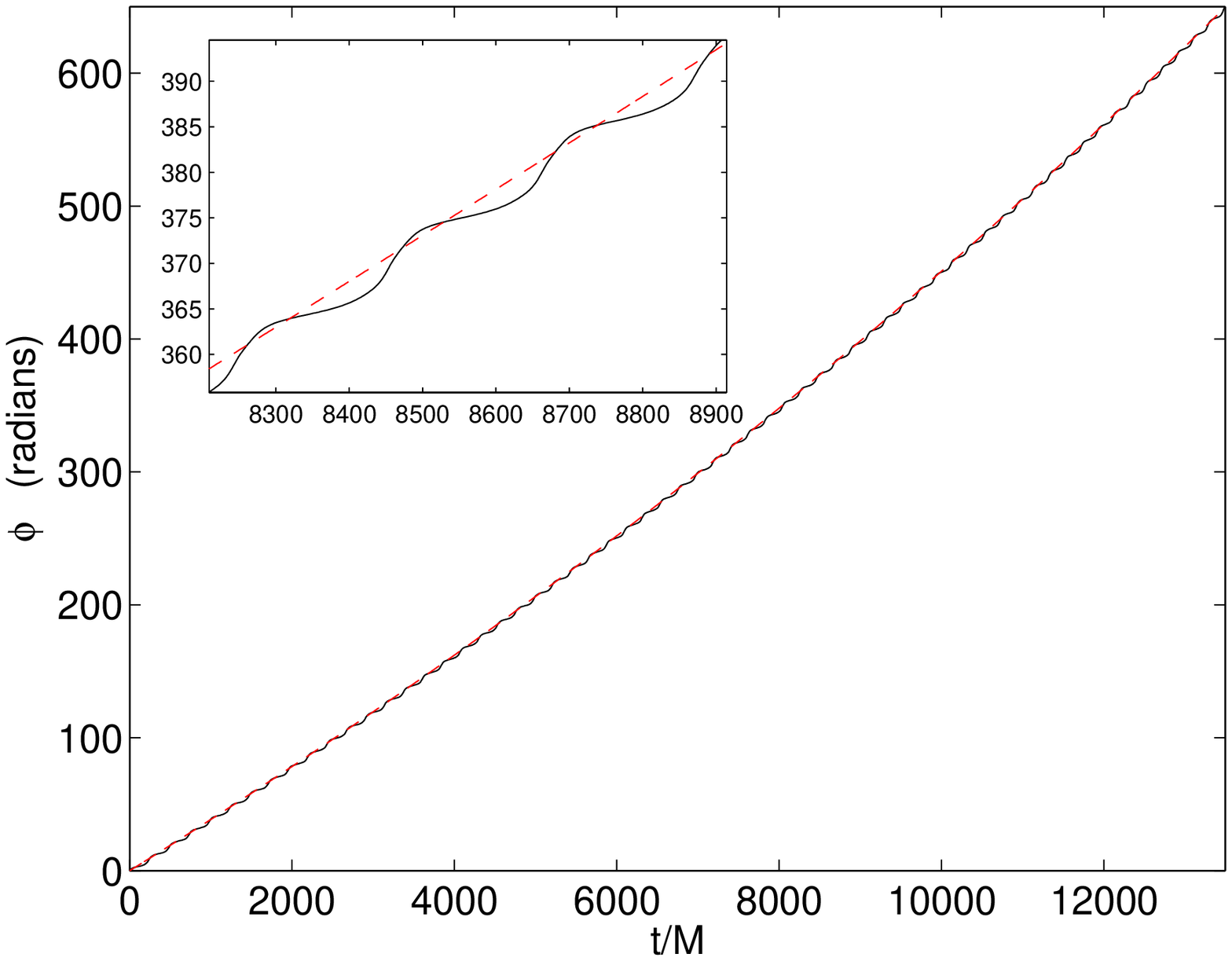}
\includegraphics[width=0.45\textwidth]{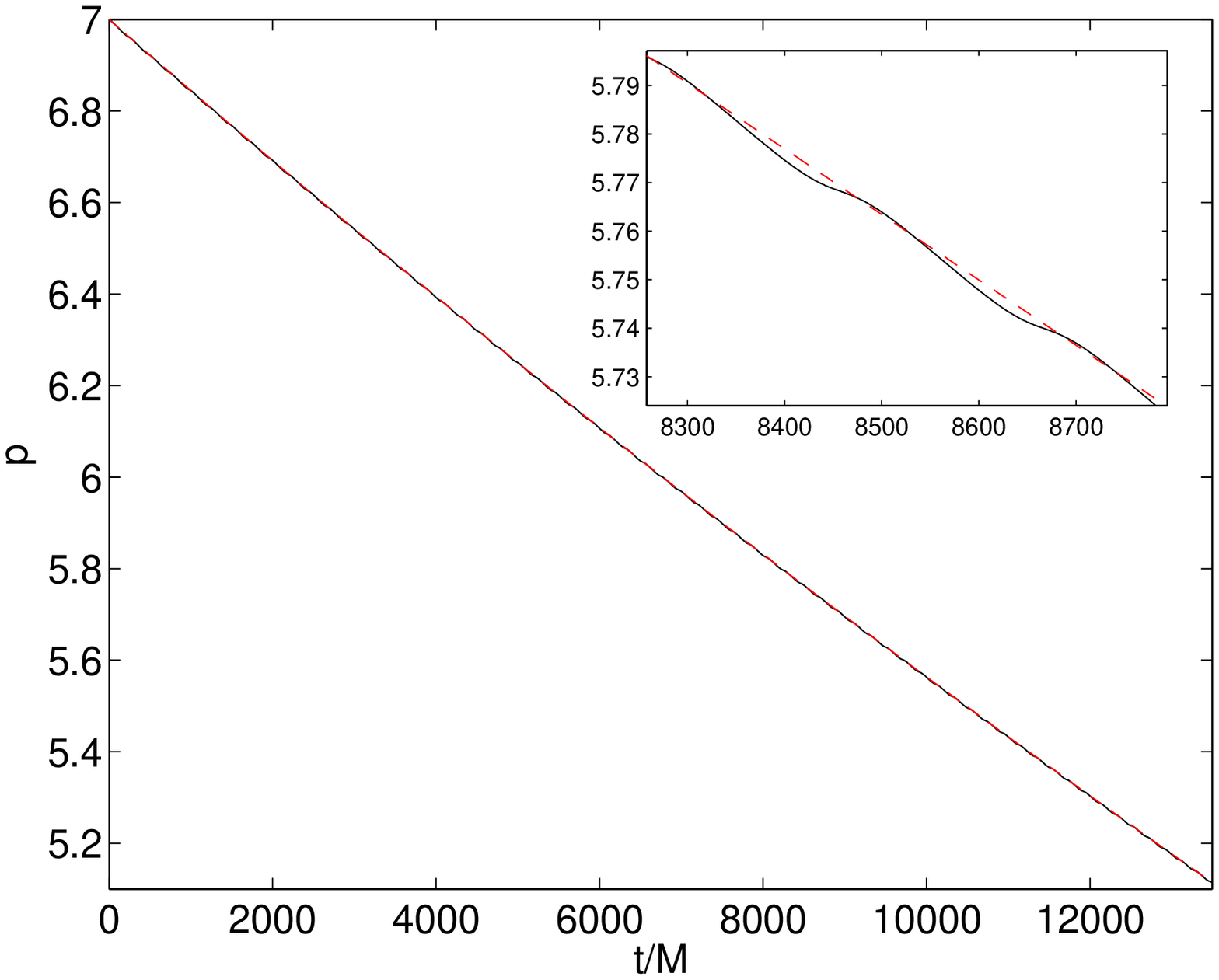}
\includegraphics[width=0.45\textwidth]{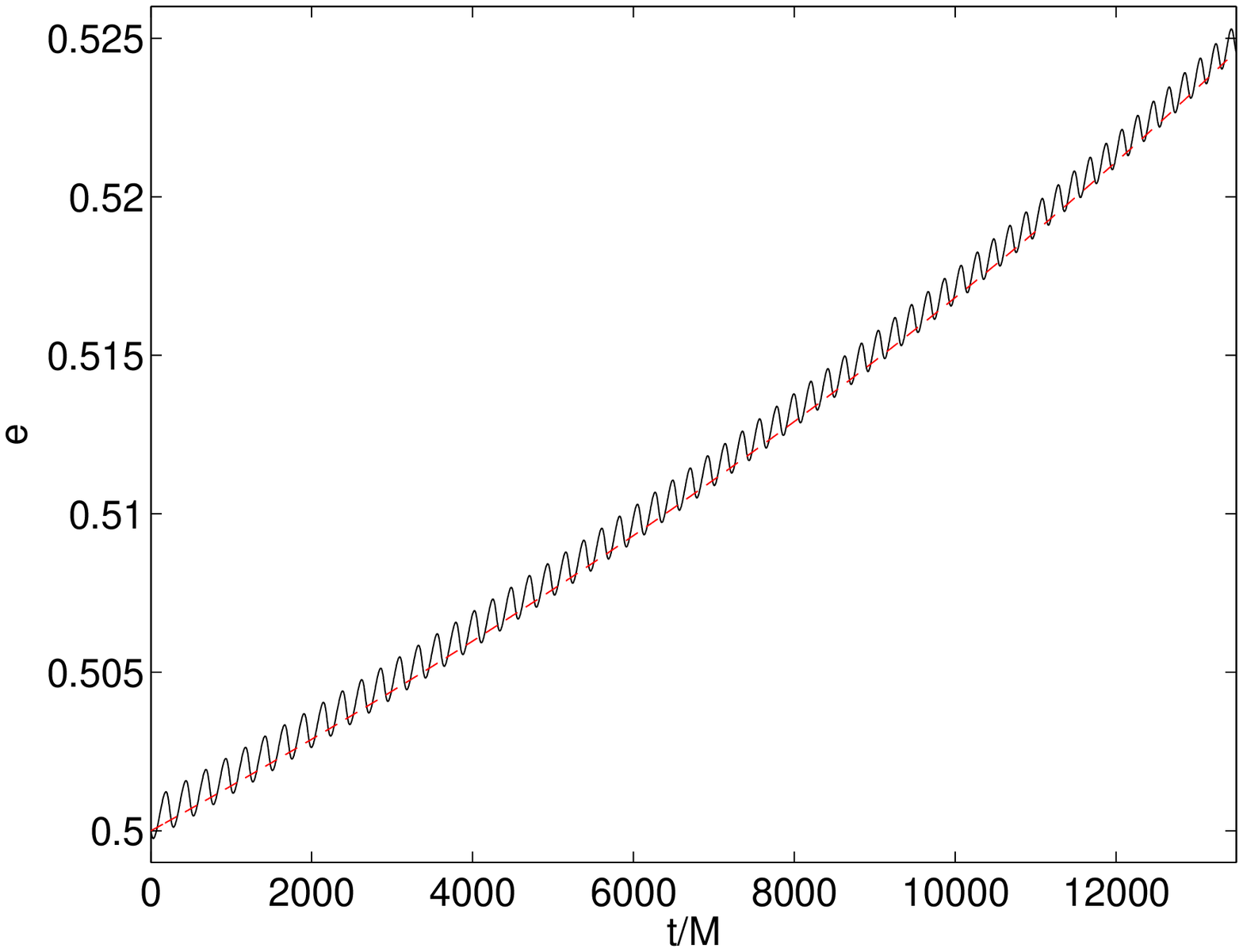}
\includegraphics[width=0.45\textwidth]{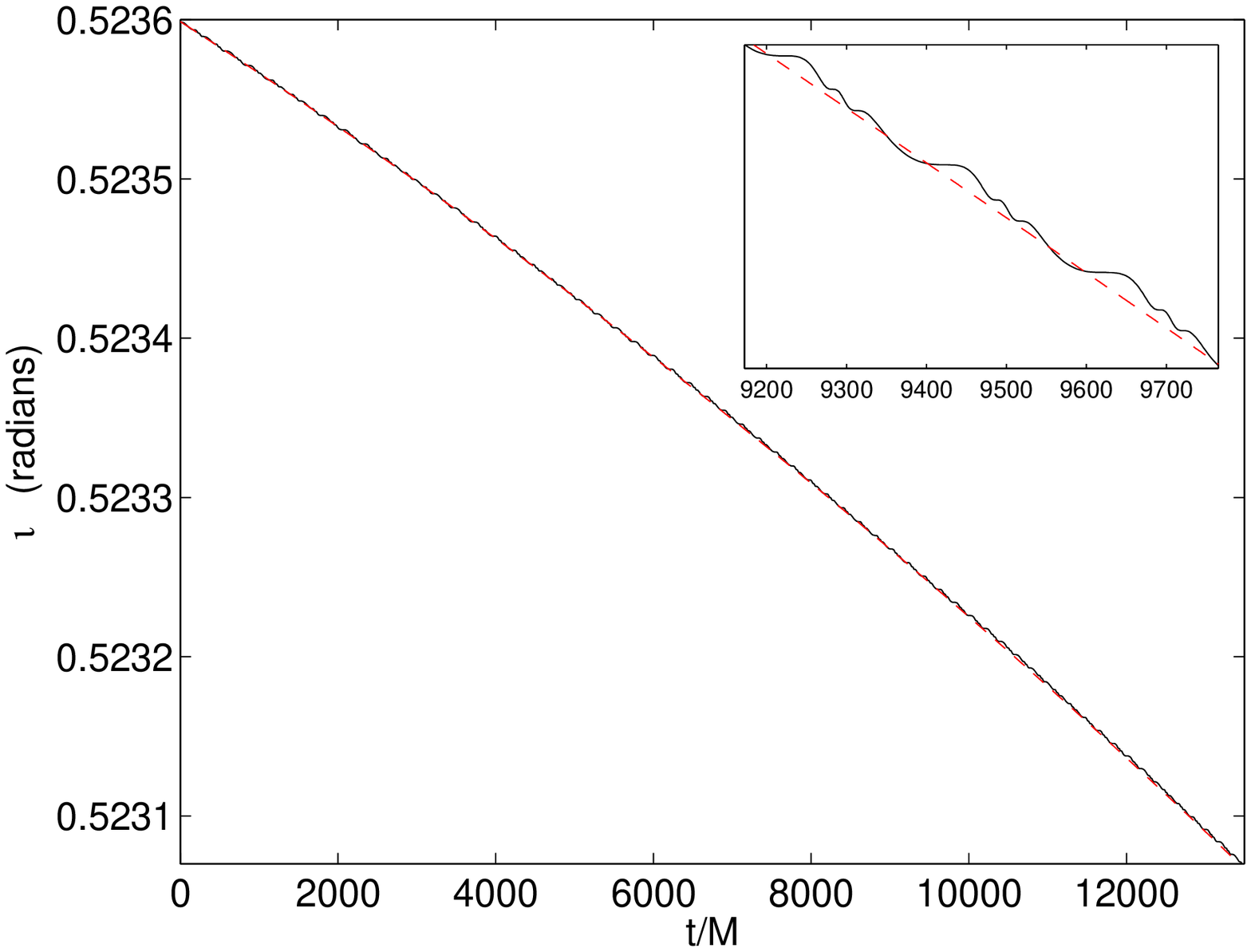}
\caption{\label{gasdragfigs}Evolution of the orbit under the influence of the ``gas-drag'' force.  The panels show, as functions of Boyer-Lindquist time $t$, the particle coordinates $r$ (top left), $\theta$ (top right) and $\phi$ (middle left) and the orbital constants $p$ (middle right), $e$ (bottom left) and $\iota$ (bottom right).
In each panel the solid curve was computed using the exact evolution, while the dashed curve was computed using the adiabatic evolution.  The plots showing $\phi$, $p$, and $\iota$ have insets which show close-up views of the same data so that the difference between the adiabatic and exact results can be seen.}
\end{figure*}

In Fig.~\ref{gasdragfigs} we show the evolution of the orbit under the influence of the gas-drag force and initial conditions given above.
The six panels show the three Boyer-Lindquist coordinates, $(r, \theta, \phi)$, and the three constants that describe the orbital shape, $(p,e,\iota)$, as functions of the Boyer-Lindquist time $t$. We see that the influence of the drag force is to drive the inspiral of the object, but also to increase the eccentricity of the orbit and decrease the orbital inclination, i.e., to make the orbit more prograde. The trajectories of the geodesic constants of motion, $(p, e, \iota)$, show oscillations on the orbital time scale, superimposed on a monotonic secular evolution on the radiation reaction time scale. The secular part is the analog of the averaged evolution, $\bar{G}^{(1)}$, described for the perturbed nonlinear oscillator in Sec.~\ref{NLosc}. The panels in Fig.~\ref{gasdragfigs} also show the solution to the adiabatic equations of motion for this problem computed using Eqs.~(\ref{generalform})--(\ref{bigxidef}). We see that the adiabatic solution is a good approximation to the average evolution along the inspiral, as expected, and, as we saw for the toy problem in Sec.~\ref{NLosc}, it provides a closer fit to the change in the constants of the motion, in this case $p$, $e$ and $\iota$, than for the phase. In this case, although the adiabatic solution remains in phase on average over the whole of the evolution seen in Fig.~\ref{gasdragfigs}, within each cycle the adiabatic solution goes in and out of phase with the true evolution. This arises because the forcing term in this case is relatively large, and so the orbit changes significantly between periapse and apoapse as a result of the forcing term. The orbit evolved using the instantaneous force therefore looks quite different from the orbit evolved continuously by the orbital-averaged force. Note that the orbits do come back into phase after each complete cycle, as expected.

This example illustrates the application of the osculating elements formalism to the computation of inspiral evolutions in the Kerr spacetime, and it serves to demonstrate that the two alternative formulations  do indeed yield the same results. However, even though the prescription for the drag force was rather simple, Fig.~\ref{gasdragfigs} also illustrates some qualitative effects of the drag force that could be used to infer the presence of such a drag from observations. For an orbit evolving under the influence of gravitational radiation reaction only, the eccentricity tends to decrease, except toward the end of the inspiral just prior to plunge~\cite{glamkenn,drascohughes06}, while the inclination tends to increase, i.e., the orbit becomes more retrograde~\cite{hughescirc1,hughescirc2,drascohughes06}. We see here that the effect of the drag force is qualitatively different, as it drives increasing eccentricity and decreasing inclination. If observed, this would provide a robust observational signature  for an orbit that was evolving under the influence of drag. A decrease in orbital inclination due to hydrodynamic drag was also seen in~\cite{BR2008}, in which a more sophisticated model for the drag force was employed. It is gratifying that this simple model produces this expected feature qualitatively. The same paper~\cite{BR2008} found that the eccentricity would increase in parts of the parameter space and decrease in other parts. For Newtonian orbits, the osculating element equations predict that the eccentricity will remain constant under the action of a simple drag force of this type (see Appendix~\ref{KepOsc}). Increasing eccentricity has, however, been seen in Newtonian simulations of binaries embedded in a realistic disc~\cite{artymowicz91,armitage05,cuadra08}. In Appendix~\ref{A:SchDrag} we show that the increase in eccentricity is an expected post-Newtonian effect and give an explanation in the context of a Schwarzschild black hole. It is clear from that discussion that this increase in eccentricity is a generic feature of relativistic drag, and so this is an observational prediction. If observed, an increasing eccentricity or decreasing inclination would be a clear signature that the observed inspiral was not occurring in a vacuum Kerr background.

\section{Discussion}
\label{S:sum}
We have described two methods for integrating the equations of motion for bound, accelerated orbits in the Kerr spacetime, which are based on identifying the orbit with a geodesic at each point. The first method parametrizes the position and velocity of the orbit in terms of the conserved quantities (energy, axial angular momentum and Carter constant) in addition to three angular variables which increase monotonically and correspond to relativistic generalizations of the anomalies of Keplerian motion.
The second method is the traditional ``osculating element'' technique which parametrizes the position and velocity of the orbit in terms of the geodesic with the same position and velocity. Practically, the second method differs from the first only in the treatment of the three phase variables, which are split up into a geodesic piece and a ``phase offset'' piece that is constant for geodesics.

To illustrate the methods, we first analyzed, as a simpler model, a forced anharmonic oscillator. This was written in terms of a set of phase space coordinates. The forced equations of motion contained an apparent divergence at the turning points, but it was possible to reformulate the equations to eliminate the problematic terms and thus obtain equations of motion in a form without divergences. We discussed the adiabatic prescription for computing the leading order motion, which corresponds to a gradual evolution of the oscillator's amplitude and fundamental frequency driven by the phase space averaged forcing function for the amplitude. We presented an  alternative analysis of this toy problem analogous to the osculating orbit method in terms of the analytic solution to the un-forced motion. By numerically integrating the equations we verified that both parametrizations gave the same results and compared these to the adiabatic approximation to the solution.

Next, we showed that the equation of forced motion in the Kerr spacetime could be reformulated in a similar fashion. For the first method, it was advantageous to parametrize the force in terms of its components on the Kinnersley tetrad instead of using the instantaneous time derivatives of the conserved quantities. We derived a formulation of the equations of motion in terms of phase variables that was manifestly divergence-free at the turning points.  We then generalized the second method, of osculating orbits, to generic orbits in the Kerr spacetime and showed how we could write down a divergent-free form of equations of this type without explicit simplification.

As an application of our results, we considered the case of a simple force that could represent a gas drag. Numerical integrations of the equations of motion for a choice of parameters verified that the two methods of parametrizing the motion gave the same results. We identified a key observational signature of the presence of a drag force, namely, a decrease in the orbital inclination and an increase in eccentricity, which is opposite to the increase in inclination and decrease in eccentricity characteristic of the gravitational radiation reaction forces during the early stage of an inspiral.

The first of our two methods has been applied to the study
of transient resonances that occur in the radiation-reaction-driven
inspirals of point particles into spinning black holes, using approximate post-Newtonian expressions for the self-force \cite{2010arXiv1009.4923F}. Other applications of this work will include the construction of accurate trajectories for orbits evolving under the action of the self-force, once self-force data for generic orbits are available. This will be essential for the construction of accurate gravitational waveforms for EMRIs, which will be needed for LISA data analysis. The formalism can also be used to estimate the magnitude of any secular changes in the orbital parameters that arise from the action of external perturbing forces. These could arise from gravitational perturbations from distant objects, such as stars or a second massive black hole, or from the presence of other material in the spacetime, such as the gas-drag which we considered in a simple way here. It will be very important to have a quantitative understanding of the importance of all these effects if intermediate-mass-ratio inspirals or EMRIs are to be used to carry out high-precision mapping of the spacetime around Kerr black holes and for tests of general relativity. Finally, the results described here will be useful to augment existing kludge models for inspiral waveforms. In particular, these methods will allow us to extract the secular part of the evolution of both the orbital constants of the motion and the phase constants, from self-force calculations. It is straightforward to include secular changes to the orbital parameters in the kludge framework~\cite{Babak:2006uv}, and by doing this it should be possible to ensure that the kludge waveform stays in phase with the true waveform for long stretches of the inspiral. It will be important to have accurate but cheap-to-calculate waveform models available when the data from gravitational wave detectors are analyzed, as this data analysis will rely heavily on matched filtering using template waveforms.

\section{Acknowledgments}
The work was supported in part by NSF Grant Nos.~PHY-0757735 and PHY-0457200 and the John and David Boochever Prize Fellowship in Theoretical Physics at Cornell and NSF Grant Nos. PHY-0653653 and PHY-0601459, CAREER Grant No. PHY-0956189, the David and Barbara Groce Start-up Fund and the Sherman Fairchild Foundation at Caltech. JG's work is supported by the Royal Society. SB and SD were supported in part by DFG Grant No.~SFB/TR 7 Gravitational Wave Astronomy and by DLR (Deutsches Zentrum fur Luft- und Raumfahrt).
EF is grateful for the hospitality of the Theoretical Astrophysics Including Relativity Group at Caltech, and the Department of Applied Mathematics and Theoretical Physics at the University of Cambridge, as this paper was being completed.

\appendix
\section{Derivation of tetrad equations of motion in terms of radial
  and polar angular variables}
\label{appendix:derivation}

In this appendix we derive the forms
Eqs.\ (\ref{eq:Eprime}) -- (\ref{eq:Kprime}), (\ref{eq:psithetaprime})
and (\ref{eq:psirprime})
of the equations of motion
for forced motion in Kerr, in the tetrad formulation, and using the
angular variables $\psi_r$ and $\psi_\theta$ instead of $r$ and
$\theta$.

\subsection{Evolution equations for first integrals $E$, $L_z$, $K$}

The evolution equations (\ref{eq:Eprime}) -- (\ref{eq:Kprime})
for the conserved quantities $E$, $L_z$ and $K$
are obtained as follows.
We start from the standard expressions for the first integrals
in terms of the Killing vectors and Killing tensors
\bea
\xi^\alpha&=&-\delta^\alpha_t~, \\
\eta^\alpha&=&\delta^\alpha_\phi~, \\
K^{\alpha\beta}&=&2\Sigma m^{(\alpha}m^{\star \beta)}-a^2\cos^2\theta g^{\alpha\beta} ~,
\label{Kofmm}
\eea
and take proper time
derivatives. Using the tetrad decomposition (\ref{eq:adecomposition})
of the acceleration together with the expressions (\ref{eq:dl}) -- (\ref{eq:dm})
for the basis covectors gives
\bea
\frac{dE}{d\tau}&=&-\vec a \cdot \frac{\partial}{\partial t} \nonumber \\
&=&-\left(a_n+\frac{\Delta}{2\Sigma} a_l\right)
-\frac{iar\sin\theta}{\sqrt{2}\Sigma}\left(a_m-a_m^\star\right) \nonumber \\
&+&\frac{a^2\sin\theta\cos\theta}{\sqrt{2}\Sigma}\left(a_m+a_m^\star\right)~.\label{eq:dedtau1}
\eea
Here, we have used the fact that ${\bf m}$ given in Eq.\ (\ref{eq:dm}) can
be written as
\bea
{\bf m}&=&\frac{1}{\sqrt{2}\Sigma}\left[-(ir+a\cos\theta)\sin\theta
dt+(r-ia\cos\theta)\Sigma d\theta \right. \nonumber \\
&+&\left. (ir+a\cos\theta)d\phi\right]~.
\eea
Noting that $a_m=(R_a-i I_a)/\sqrt{2}$, and eliminating $a_l$ with
the aid of Eq.\ (\ref{eq:alexpr}) transforms Eq.\ (\ref{eq:dedtau1})
to the form
\bea
 \frac{dE}{d\tau}&=&-\frac{a_n}{u_n}\left(u_n-\frac{\Delta}{2\Sigma}
 u_l\right)-\frac{\Delta}{2\Sigma
 u_n}\left(R_aR_u+I_aI_u\right) \nonumber \\
 &-&\frac{a\sin\theta}{\Sigma}\left(rI_a-a\cos\theta R_a\right)
\label{eq:dedtau2}~.
\eea
Using the expressions (\ref{eq:velcomponents})
for the tetrad components $u_n$ and $u_l$ of the four velocity
and converting
from $\tau$ derivatives to $\lambda$ derivatives using Eq.\
(\ref{eq:Minotime}) then leads to the final form given in Eq.\
(\ref{eq:Eprime}).

Similarly, we obtain Eq.\ (\ref{eq:Lzprime}) by starting from
\bea
\frac{dL_z}{d\tau}&=&\vec a \cdot \frac{\partial}{\partial \phi} \nonumber \\
&=& -a\sin^2\theta\left(a_n+\frac{\Delta}{2\Sigma}a_l\right)-\frac{\varpi^2
r\sin\theta}{\Sigma}I_a \nonumber \\
&+&\frac{a\varpi^2\sin\theta\cos\theta}{\Sigma}R_a~,
\eea
and eliminating $a_l$ using Eq.\ (\ref{eq:alexpr}) to obtain
\bea
\frac{dL_z}{d\tau}&=&-a\sin^2\theta
\frac{a_n}{u_n}\left(u_n+\frac{\Delta}{2\Sigma}u_l\right) \nonumber \\
&-&\frac{a\sin^2\theta}{2\Sigma u_n}\left(R_aR_u+I_a I_u\right)\nonumber \\
&-&\frac{\varpi^2 r\sin\theta}{\Sigma}I_a+\frac{a\varpi^2\sin\theta\cos\theta}{\Sigma} R_a~.
\eea
The form quoted in Eq.\ (\ref{eq:Lzprime}) is then obtained
from this using Eqs.\ (\ref{eq:velcomponents}) as before.

The evolution of the Carter constant is obtained very simply from
the expression for the Killing tensor to be
\be
\frac{dK}{d\tau}=2K^{\alpha\beta}u_\alpha a_\beta=2\Sigma \left(R_uR_a+I_uI_a\right)~,
\label{dKdt}
\ee
where we have used the orthogonality relation
$g^{\alpha\beta}u_\alpha a_\beta=0$ and written the combination
$(u_m^\star a_m+u_m a_m^\star)$ in terms of $R_u$ and $R_a$.
Combining this with the definition (\ref{eq:calAIII}) of ${\cal
  A}_{III}$ yields Eq.\ (\ref{eq:Kprime}).


\subsection{Polar motion}

To obtain the equation of motion (\ref{eq:psithetaprime}) for
$\psi_\theta$, we start by
differentiating its definition (\ref{eq:psithetadef}) with respect
to $\lambda$:
\bea
 \sin\theta\cos\theta
 \left(\frac{d\theta}{d\lambda}\right)&=&z_-\sin\psi_\theta\cos\psi_\theta \nonumber \\
&\times& \left[\left(\frac{d\psi_\theta}{d\lambda}\right)-\frac{1}{2z_-}\cot\psi_\theta
   \left(\frac{dz_-}{d\lambda}\right)\right]~.\nonumber \\ && \label{eq:dp1}
\eea
The equation of motion (\ref{eq:thetadot}) for $\theta$ can
be rewritten in the form \be
\left(\frac{d\theta}{d\lambda}\right)^2=\beta  z_-\sin^2\psi_\theta
\frac{\left(z_+-z_-\cos^2\psi_\theta\right)}{\left(1-z_-\cos^2\psi_\theta\right)}~.\label{eq:dp2}
\ee
%
%
We now take the square root of this equation.
From the definition (\ref{eq:psithetadef}) of $\psi_\theta$ and
noting that $\psi_\theta$ monotonically
increases, we see that $(d\theta/d\lambda)>0$ for
$0<\psi_\theta<\pi$ and $(d\theta/d\lambda)<0$ for
 $\pi<\psi_\theta<2\pi$, so we must choose the positive square
root on the right-hand side.
Combining Eq.\ (\ref{eq:dp1}) and the square root of Eq.\
(\ref{eq:dp2}) now leads to an
equation of motion for $\psi_\theta$ in the form
\be
\frac{d\psi_\theta}{d\lambda}=\sqrt{\beta
  (z_+-z_-\cos^2\psi_\theta)}+\frac{\cot\psi_\theta}{2z_-}\frac{dz_-}{d\lambda}\label{eq:dp3}.
\ee

Next, we can obtain an expression for $dz_-/d\lambda$ in terms of
$dP_i/d\tau$, where $P_i=(E,L_z,K)$ are the constants of motion, as
follows.  Using the chain rule gives $dz_-/d\lambda=(\partial z_-/\partial
P_i)(dP_i/d\lambda)$. Differentiating $V_\theta$ in the
form given in Eq.\ (\ref{eq:vthetaofz}) with respect to $P_i$ at
fixed $z$ and evaluating the result at $z_-$ relates $dz_-/dP_i$ to
$(\partial V_\theta/\partial P_i)_{z_-}$, which can be computed from
Eq.\ (\ref{eq:vthetaz1}). This yields
\bea
\frac{\beta(z_+-z_-)}{(1-z_-)}\frac{dz_-}{d\lambda}&=&
\frac{dQ}{d\lambda}-2 L_z
\left(\frac{z_-}{1-z_-}\right)\frac{dL_z}{d\lambda} \nonumber \\
&+&2a^2E z_-\frac{dE}{d\lambda}~.
\eea
Now switching from the Carter constant $Q$
to $K=Q+(L_z-aE)^2$ and using that $d\lambda =d\tau/\Sigma$ we
obtain
\bea
\frac{\beta(z_+-z_-)}{\Sigma}\frac{dz_-}{d\lambda}&=&(1-z_-)\frac{dK}{d\tau}-2
\left(L_z-a(1-z_-)E\right) \nonumber \\
&\times&\left[\frac{dL_z}{d\tau}-a(1-z_-)\frac{dE}{d\tau}\right].\label{eq:zminusevol}
\eea

The expressions for the evolution of $P_i$ in
Eqs.\ (\ref{eq:Eprime})  -- (\ref{eq:Kprime}) can now be used to obtain the
explicit dependence on $\psi_\theta$ of  $dz_-/d\lambda$ given by
Eq.\ (\ref{eq:zminusevol}) by direct substitution. This gives
\begin{widetext}
\bea
\beta (z_+-z_-)\frac{dz_-}{d\lambda}&=&\frac{2\Delta {\cal H}_{-}u_r
  a_n}{u_n} \ a z_-
\sin^2\psi_\theta+\left[2(1-z_-)(r^2+a^2z_-\cos^2\psi_\theta)^2-az_-{\cal
    H}_{-}\frac{\Delta
    \sin^2\psi_\theta}{u_n}\right]\left(R_uR_a+I_uI_a\right)\nonumber\\
&&+2{\cal
  H}_-\sqrt{1-z_-\cos^2\psi_\theta}(r^2+a^2
  z_-)\left(rI_a-a\cos\theta R_a\right)~,
\eea
where ${\cal H}_-={\cal H}(z_-)=L_z-a(1-z_-)E$. This can be written as
\bea
\beta (z_+-z_-)\frac{dz_-}{d\lambda}&=&\frac{\Delta {\cal H}_{-}}{u_n} \ a z_-
\sin^2\psi_\theta\left[(R_uR_a+I_uI_a)-2 u_r a_n\right]\nonumber\\
&+&2\left[(1-z_-)(r^2+a^2z_-\cos^2\psi_\theta)^2R_u-{\cal H}_-a\sqrt{z_-}(r^2+a^2z_-)
  \sqrt{1-z_-^2\cos^2\psi_\theta}
  \ \cos\psi_\theta\right]R_a\nonumber\\
&+&2\left[(1-z_-)(r^2+a^2z_-\cos^2\psi_\theta)^2I_u+{\cal H}_- r (r^2+a^2z_-)
  \sqrt{1-z_-^2\cos^2\psi_\theta}\right]I_a~.
\eea
Substituting the expressions for the four velocity components
Eqs.\ (\ref{eq:velcomponents}) and the definition (\ref{eq:calHdef}) of ${\cal H}$ into
the coefficients of $R_a$ and $I_a$ inside the square brackets and
expanding them out gives
\bea
\beta (z_+-z_-)\frac{dz_-}{d\lambda}&=&\frac{\Delta {\cal H}_{-}}{u_n} \ a z_-
\sin^2\psi_\theta\left[{\cal A}_{III}-2 u_r a_n\right]+2(1-z_-)\Sigma u_\theta (a \cos\theta I_a+rR_a)\nonumber\\
&+&\frac{2rz_-\sin^2\psi_\theta}{\sqrt{1-z_-\cos^2\psi_\theta}}\left[\varpi^3
  L_z-a^3E(1-z_-)(1-z_-\cos^2\psi_\theta)\right]
(rI_a-a\cos\theta R_a)~.\label{eq:dzminuslast}
\eea
Using that $u_\theta=(d\theta/d\lambda)$, with $(d\theta/d\lambda)$
given by the positive square root of Eq.\ (\ref{eq:dp2}) and inserting
Eq.\ (\ref{eq:dzminuslast}) into the equation of motion for
$\psi_\theta$ of Eq.\ (\ref{eq:dp3}) leads to the final result quoted
in Eq.\ (\ref{eq:psithetaprime}).
\end{widetext}

\subsection{Radial motion}

We now give a derivation of the radial equation of motion
(\ref{eq:psirprime}) which is similar to the above derivation of the
equation (\ref{eq:psithetaprime}) of polar motion.
From the definitions (\ref{eq:rdot}) and (\ref{Vrdef}) of the radial potential we have
\bea
\left(\frac{dr}{d\lambda}\right)^2&=&F^2-\Delta (r^2+K)\nonumber \\
&=&(1-E^2)(r_1-r)(r-r_2)(r-r_3)(r-r_4)~,\nonumber \\ \label{eq:rmotionapp}
\eea
where $F$ was defined in Eq.\ (\ref{eq:Fdef}).  We parametrize the
roots of the right-hand side by Eq.\ (\ref{eq:rturningptdef}) for the
turning points $r_1$ and $r_2$ of the bound motion, and by
\be
 r_3=\frac{p_3}{1-e}~, \ \ \ \ \ r_4=\frac{p_4}{1+e}~\label{eq:r3r4def}
\ee
for the other two roots.
Substituting the definition (\ref{eq:psirdef})
of $\psi_r$ into Eq.\ (\ref{eq:rmotionapp})
and using Eqs.\ (\ref{eq:rturningptdef}) and  (\ref{eq:r3r4def}) gives,
after some algebra,
\bea
\left(\frac{dr}{d\lambda}\right)^2&=&\frac{(1-E^2)p^2e^2\sin^2\psi_r}{(1-e^2)^2(1+e\cos\psi_r)^4}
\nonumber \\
&\times& \left[p(1-e)-p_3(1+e\cos\psi_r)\right]
\nonumber \\
&\times& \left[p(1+e)-p_4(1+e\cos\psi_r)\right]~.\label{eq:vrapp}
\eea
By differentiating the definition
(\ref{eq:psirdef}) of $\psi_r$ we obtain
\bea
\frac{d\psi_r}{d\lambda}&=&\frac{(1+e\cos\psi_r)^2}{ep\sin\psi_r}
\ \left(\frac{dr}{d\lambda}\right)+\frac{\cot\psi_r}{e}\left(\frac{de}{d\lambda}\right)\nonumber \\
&-&\frac{1+e\cos\psi_r}{ep\sin\psi_r}\left(\frac{dp}{d\lambda}\right)~.\label{eq:dpsir}
\eea
We note that $\psi_r$ is chosen to monotonically increase, which means
$d\psi_r/d\lambda>0$. We specialize to the convention that $\psi_r=0$
at $r=r_2$ and $\psi_r=\pi$ at $r=r_1$, so that $r$ increases for
$0<\psi_r<\pi$ and decreases for $\pi<\psi_r<2\pi$ and we
choose the positive square root in Eq.\ (\ref{eq:vrapp}).
Substituting Eq.\ (\ref{eq:vrapp}) for $dr/d\lambda$ in
Eq.\ (\ref{eq:dpsir}) shows that the geodesic term
becomes
\bea
\left.\frac{d\psi_r}{d\lambda}\right\rvert_{\rm geodesic}&=&
\frac{\sqrt{1-E^2}}{(1-e^2)}\left[p(1-e)-p_3(1+e\cos\psi_r)\right]^{1/2} \nonumber \\
&\times& \left[p(1+e)-p_4(1+e\cos\psi_r)\right]^{1/2} \nonumber \\
&=&{\cal P}~.\label{eq:calPus}
\eea
Here one can check that Eq.\ (\ref{eq:calPus}) is just a
reparametrization of Eq.\ (\ref{eq:def_j}) by substituting the radial
potential in the form given in Eq.\ (\ref{eq:rdot}) in terms of $P_i$ into
Eq.\ (\ref{eq:dpsir}), since ${\cal P}=(dr/d\psi_r)^{-1}\sqrt{V_r}$,
expressed in terms of $\psi_r$.

The nongeodesic terms in Eq.\ (\ref{eq:dpsir}) are obtained as
follows. From Eq.\ (\ref{eq:rturningptdef}) for $r_1$ and $r_2$ it
follows that $2p^{-1}=r_1^{-1}+r_2^{-1}$ and $2(1-e)^{-1}=r_1/r_2+1$, and
thus
\bes
\label{eq:pnedot}
\bea
2\frac{dp}{d\lambda}&=&p^2\left(\frac{dr_1/d\lambda}{r_1^2}+\frac{dr_2/d\lambda}{r_2^2}\right)
\nonumber\\
&=&(1-e)^2\frac{dr_1}{d\lambda}+(1+e)^2\frac{dr_2}{d\lambda}~,\\
2\frac{de}{d\lambda}&=&
p^2\left(\frac{dr_1/d\lambda}{r_1^2r_2}-\frac{dr_2/d\lambda}{r_2^2r_1}\right)
\nonumber \\
&=&\frac{(1-e^2)}{p}\left[(1-e)\frac{dr_1}{d\lambda}-(1+e)\frac{dr_2}{d\lambda}\right]~.
\nonumber \\&&
\eea
\ees
Substituting Eqs.~(\ref{eq:pnedot}) into Eq.\ (\ref{eq:dpsir})  gives
\bea
\frac{d\psi_r}{d\lambda}&=&{\cal P}+\frac{1}{2ep\sin\psi_r}\nonumber\\
&\times& \left[(1-e)^2(\cos\psi_r-1)\frac{dr_1}{d\lambda}\right. \nonumber\\
&&-\left. (1+e)^2(1+\cos\psi_r)\frac{dr_2}{d\lambda}\right]~.
\eea

Next, expressions for the derivatives of the turning points $r_1$ and $r_2$ can be computed
in terms of
$dP_i/d\lambda$ by using that
$(dr_{1,2}/d\lambda)=(\partial r_{1,2}/\partial P_i)dP_i/d\lambda$.
Differentiating the radial potential with respect to $P_i$ at fixed $r$ and
evaluating the result at $r_1$ and $r_2$ gives
\bea
\left.\frac{\partial V_r}{\partial P_i}\right\rvert_{r_1}&=&
(1-E^2)(r_1-r_2)(r_1-r_3)(r_1-r_4)\frac{\partial r_1}{\partial P_i}~, \nonumber\\\label{eq:dvrdpi1}&&\\
\left.\frac{\partial V_r}{\partial P_i}\right\rvert_{r_2}&=&-(1-E^2)(r_1-r_2)(r_2-r_3)(r_2-r_4)\frac{\partial r_2}{\partial P_i}~.\nonumber\\&& \label{eq:dvrdpi}
\eea
We note that one can see from Eqs.~(\ref{Vrdef}), (\ref{eq:rmotionapp}) and (\ref{eq:dvrdpi1})--(\ref{eq:dvrdpi})
that the coefficients of $\partial
r_{1,2}/\partial P_i$ can be
expressed in terms of the $r$-derivative of $V_r$ at fixed $P_i$ evaluated at the
turning points as
\bea
\left.\frac{\partial V_r}{\partial P_i}\right\rvert_{r_{1,2}}&=&-\left.\frac{\partial
  V_r}{\partial r}\right\rvert_{r_{1,2}}~\frac{\partial r_{1,2}}{\partial P_i}, \\
&=&-\kappa(r_{1,2})\frac{\partial r_{1,2}}{\partial P_i}~.
\eea
Here $\kappa(r) \equiv V_r'(r)$, which can be computed from
Eq.\ (\ref{eq:rmotionapp}) to be
\be
\kappa(r)=4EFr-2r\Delta-2(r-M)(r^2+K)\label{eq:kapparesult}~,
\ee
where the definition (\ref{eq:Fdef}) of $F$ has been used.
Using the derivatives of Eq.\ (\ref{eq:rmotionapp}) with respect to
$P_i$ then results in the following expressions for
$dr_{1,2}/d\lambda$:
\be
\frac{dr_{1,2}}{d\lambda}=-\frac{2F_{1,2}}{\kappa_{1,2}}\left(\varpi^2_{1,2}\frac{dE}{d\lambda}-a\frac{dL_z}{d\lambda}\right)+\frac{\Delta_{1,2}}{\kappa_{1,2}}\frac{dK}{d\lambda}\label{eq:r12prime}~.
\ee
With this, Eq.\ (\ref{eq:dpsir}) becomes
\begin{widetext}
\bea
\frac{d\psi_r}{d\lambda}={\cal
  P}&+&\frac{1}{2ep\sin\psi_r}\left\{(1-e)^2(\cos\psi_r-1)\left[-\frac{2F_1}{\kappa_1}\left(\varpi^2_1\frac{dE}{d\lambda}-a\frac{dL_z}{d\lambda}\right)+\frac{\Delta_1}{\kappa_1}\frac{dK}{d\lambda}\right]\right.\nonumber\\
&& \ \ \ \ \ \left.-(1+e)^2(\cos\psi_r+1)\left[-\frac{2F_2}{\kappa_2}\left(\varpi^2_2\frac{dE}{d\lambda}-a\frac{dL_z}{d\lambda}\right)+\frac{\Delta_2}{\kappa_2}\frac{dK}{d\lambda}\right]\right\}~.\label{eq:dpsir1}
\eea
The next step is to substitute the expressions
(\ref{eq:Eprime}) -- (\ref{eq:Kprime}) for the derivatives
of the first integrals into
Eq.\ (\ref{eq:dpsir1}). After some algebra we obtain
\bea
2ep\sin\psi_r\left(\frac{d\psi_r}{d\lambda}-{\cal P}\right)&=& 2\Delta
\frac{u_r}{u_n}a_n\left[(1-e)^2(1-\cos\psi_r)\frac{\Sigma_1F_1}{\kappa_1}+(1+e)^2(1+\cos\psi_r)\frac{\Sigma_2F_2}{\kappa_2}\right]\nonumber\\
&+&R_a(1-e)^2(1-\cos\psi_r)\left[R_u\left(\frac{\Sigma_1F_1\Delta}{\Sigma\kappa_1u_n}+\frac{2\Sigma\Delta_1}{\kappa_1}\right)+\frac{2F_1a^2\sin\theta\cos\theta(r^2-r_1^2)}{\kappa_1\Sigma}\right]\nonumber\\
&+&I_a(1-e)^2(1-\cos\psi_r)\left[I_u\left(\frac{\Sigma_1F_1\Delta}{\Sigma\kappa_1u_n}+\frac{2\Sigma\Delta_1}{\kappa_1}\right)-\frac{2F_1a
    r\sin\theta(r^2-r_1^2)}{\kappa_1\Sigma}\right]\nonumber\\
&+&R_a(1+e)^2(1+\cos\psi_r)\left[(1\leftrightarrow
  2)\right]+I_a(1+e)^2(1+\cos\psi_r)\left[(1\leftrightarrow
  2)\right]~,\label{eq:dpsirmess}
\eea
\end{widetext}
where $\Sigma_1=\varpi^2_1-a^2 \sin^2\theta$. Noting that $u_r=\Delta^{-1}(dr/d\lambda)$ and using the definition (\ref{eq:calPus}) of
${\cal P}$ gives an explicit expression for
$u_r$:
\be
u_r=\frac{pe\sin\psi_r {\cal P}}{\Delta (1+e\cos\psi_r)^2}~.\label{eq:urexplicit}
\ee
Also, from the definitions (\ref{eq:psirdef}) and (\ref{eq:rturningptdef}), we have that
\bea
(r-r_1)&=&-\frac{pe(1+\cos\psi_r)}{(1-e)(1+e\cos\psi_r)}~, \\
(r-r_2)&=&\frac{pe(1-\cos\psi_r)}{(1+e)(1+e\cos\psi_r)}~.\label{eq:rr12}
\eea
Substitution of Eq.\ (\ref{eq:urexplicit}) and Eq.\ (\ref{eq:velcomponents})
together with further algebraic manipulations on Eq.\ (\ref{eq:dpsirmess}) lead to
\begin{widetext}
\bea
\frac{d\psi_r}{d\lambda}&=& {\cal
  P}\left\{1+\frac{a_n}{u_n(1+e\cos\psi_r)^2}\left[(1-e)^2(1-\cos\psi_r)\frac{\Sigma_1F_1}{\kappa_1}+(1+e)^2(1+\cos\psi_r)\frac{\Sigma_2F_2}{\kappa_2}\right]\right\}\nonumber\\
&&+\frac{(1-e)^2(\cos\psi_r-1)}{2ep\sin\psi_r} \left[\frac{1}{\kappa_1
    u_n}\left(\Sigma_1F_1\Delta-\Sigma\Delta_1F\right)-\frac{\Delta\Sigma\Delta_1u_r}{\kappa_1u_n}\right](R_uR_a+I_uI_a)\nonumber\\
&&+\frac{(1-e)^2(\cos\psi_r-1)F_1a\sin\theta}{\kappa_1ep\sin\psi_r}(r+r_1)(r-r_1)(a\cos\theta
R_a-rI_a)\nonumber\\
&&+\frac{(1+e)^2(1+\cos\psi_r)}{2ep\sin\psi_r} \left[\frac{1}{\kappa_2
    u_n}\left(\Sigma_2F_2\Delta-\Sigma\Delta_2F\right)-\frac{\Delta\Sigma\Delta_2u_r}{\kappa_2u_n}\right](R_uR_a+I_uI_a)\nonumber\\
&&+\frac{F_2a\sin\theta(1+e)^2(1+\cos\psi_r)}{\kappa_2ep\sin\psi_r}(r+r_2)(r-r_2)(a\cos\theta
R_a-rI_a)~.\label{eq:psirmess1}
\eea
We can  simplify the coefficients of $R_a$ and $I_a$ by expanding the term
$(\Sigma_1F_1\Delta-\Sigma\Delta_1 F)$ using the explicit expressions
in Eq.\ (\ref{eq:sigmaexpr}) to obtain an explicit factor of $(r-r_1)$:
\bea
\frac{(\Sigma_1F_1\Delta-\Sigma\Delta_1 F)}{(r-r_1)}&=&-(r+r_1)\left[a^3 (a E+ z L_z)+E \left(r^2+r_1^2\right)
   \left(a^2-2 M r\right)-2 a^2 E M r z-4 a^2 E M r+2 a
  L_z M r+E r^2r_1^2\right]\nonumber \\
&\equiv& Q_1~,\label{eq:q1appendix}
\eea
where $z=\cos^2\theta$, as before. We similarly define $Q_2$ by
replacing $1\to 2$ in Eq.\ (\ref{eq:q1appendix}). Substituting
Eqs.\ (\ref{eq:q1appendix}) as well as Eqs.\ (\ref{eq:urexplicit}) and
(\ref{eq:rr12}) into Eq.\ (\ref{eq:psirmess1}) and using the
definitions (\ref{eq:calAdef}) yields after simplifications
\bea
\frac{d\psi_r}{d\lambda}&=& {\cal
  P}\left\{1+\frac{a_n}{u_n(1+e\cos\psi_r)^2}\left[(1-e)^2(1-\cos\psi_r)\frac{\Sigma_1F_1}{\kappa_1}+(1+e)^2(1+\cos\psi_r)\frac{\Sigma_2F_2}{\kappa_2}\right]\right\}\nonumber\\
&&+\frac{(1-e)^2(1-\cos\psi_r)}{2\sin\psi_r}\left\{\frac{\Sigma\Delta_1
  {\cal P}{\cal A}_{III}\sin\psi_r}{\kappa_1
  u_n(1+e\cos\psi_r)^2}+\frac{Q_1 {\cal
    A}_{III}(1+\cos\psi_r)}{(1+e\cos\psi_r)(1-e)}-\frac{2F_1
  a\sin\theta (r+r_1){\cal
    A}_{II}(1+\cos\psi_r)}{\kappa_1(1-e)(1+e\cos\psi_r)}\right\}\nonumber\\
&&+\frac{(1+e)^2(1+\cos\psi_r)}{2\sin\psi_r}\left\{\frac{\Sigma\Delta_2
  {\cal P}{\cal A}_{III}\sin\psi_r}{\kappa_2
  u_n(1+e\cos\psi_r)^2}-\frac{Q_2 {\cal
    A}_{III}(1-\cos\psi_r)}{(1+e\cos\psi_r)(1+e)}+\frac{2F_2
  a\sin\theta (r+r_2){\cal
    A}_{II}(1-\cos\psi_r)}{\kappa_2(1+e)(1+e\cos\psi_r)}\right\}~.\nonumber
\eea
This can be further simplified to be
\bea
\frac{d\psi_r}{d\lambda}&=& {\cal
  P}\left\{1+\frac{a_n}{u_n(1+e\cos\psi_r)^2}\left[(1-e)^2(1-\cos\psi_r)\frac{\Sigma_1F_1}{\kappa_1}+(1+e)^2(1+\cos\psi_r)\frac{\Sigma_2F_2}{\kappa_2}\right]\right\}\nonumber\\
&&+\frac{{\cal
    A}_{III}\sin\psi_r}{2(1+e\cos\psi_r)u_n}\left[\frac{Q_1(1-e)}{\kappa_1}-\frac{Q_2(1+e)}{\kappa_2}\right]\nonumber\\
&&+\frac{\Sigma
  {\cal A}_{III}{\cal P}}{2(1+e\cos\psi_r)^2
  u_n}\left[(1-e^2)(1-\cos\psi_r)\frac{\Delta_1}{\kappa_1}+(1+e)^2(1+\cos\psi_r)\frac{\Delta_2}{\kappa_2}\right]\nonumber\\
&&-\frac{a\sin\theta\sin\psi_r{\cal A}_{II}}{1+e\cos\psi_r}\left[\frac{F_1(1-e)(r+r_1)}{\kappa_1}-\frac{F_2(1+e)(r+r_2)}{\kappa_2}\right]~.
\eea
\end{widetext}

\section{Adiabatic Limit}
\label{app:adiabatic}

In this appendix, we derive our method of obtaining the leading order,
adiabatic solutions to the forced geodesic equations in Kerr.
This method was used to obtain the numerical adiabatic solutions that
are plotted and discussed in Sec.\ \ref{gasdrag} above.
The starting point is
the specific form (\ref{eq:Eprime}) -- (\ref{eq:psirprime}) of the
forced geodesic equations derived in
Sec.\ \ref{tetform} above, which
have the general form
\bes
\label{gs}
\bea
{\dot \psi}_\alpha &=& \omega_{\alpha}(\psi_\alpha, {\bf J}) +
\epsilon g^{(1)}_\alpha(\bfpsi, {\bf J}) + O(\epsilon^2), \nonumber\\&&\hspace{1.4in}~1 \le \alpha \le N,
\\
{\dot J}_\lambda &=& \epsilon G^{(1)}_\lambda(\bfpsi, {\bf J}) +
\epsilon^2 G^{(2)}_\lambda(\bfpsi, {\bf J}) + O(\epsilon^3),\nonumber\\&&\hspace{1.4in}~ 1 \le \lambda \le M.
\eea
\ees
Here $\bfpsi = (\psi_1, \ldots, \psi_N)$ are a set of angular
variables, and ${\bf J} = (J_1, \ldots, J_M)$ are a set of quantities that
are conserved for the unperturbed system.  Dots denote derivatives
with respect to $\lambda$.  The functions
$\omega_\alpha$ determine the frequencies of the unperturbed motion
(geodesic motion for the Kerr application),
and the functions $g^{(1)}_\alpha$, $G^{(1)}_\lambda$ and
$G^{(2)}_\lambda$ represent the external perturbations on the
system\footnote{Note that the notation $\omega_\alpha(\psi_\alpha,{\bf
    J})$ means that
each $\omega_\alpha$ depends only on a single phase variable
$\psi_\alpha$, and does not depend on the phase variables $\psi_\beta$
with $\beta\ne\alpha$.  The adiabatic limit of the more general system of equations
with $\omega_\alpha = \omega_\alpha(\bfpsi,{\bf J})$ would be
considerably more complicated.}.
These functions are all periodic in each phase variable with period $2 \pi$.
In the special case when the frequencies $\omega_\alpha$ are
independent of the phase variables $\bfpsi$, the variables
$\psi_\alpha$ and $J_\lambda$ are (generalized versions of) action-angle variables.  This special case is actually fully general; one can
always perform a redefinition of the phase variables to achieve this.
This case of action-angle variables was studied in detail in Ref.\
\cite{FH}, where the form of the adiabatic and post-adiabatic
solutions were derived.

Here we will generalize the analysis of Ref.\ \cite{FH} to the more
general system of Eqs.~(\ref{gs}), since our system of Eqs.~(\ref{eq:Eprime})--(\ref{eq:psirprime}) in Kerr is of this form.
We start by describing the result for the adiabatic limit, and then we
outline its derivation.  The adiabatic solutions are given by the
following set of steps:
\begin{enumerate}
\item We define the averaging operation, for any function $f(\bfpsi)$ of
  $\bfpsi$, by
\be
\left< f \right>_{\bf J} \equiv \frac{ \int_0^{2 \pi}
\frac{ d\psi_1}{\omega_1(\psi_1,{\bf J})} \ldots \int_0^{2 \pi} \frac{
  d \psi_N}{\omega_N(\psi_N,{\bf J})} f(\psi_1, \ldots, \psi_N)}{
\int_0^{2 \pi}
\frac{ d\psi_1}{\omega_1(\psi_1,{\bf J})} \ldots \int_0^{2 \pi} \frac{
  d \psi_N}{\omega_N(\psi_N,{\bf J})} }~.
\label{averagedef}
\ee
The subscript ${\bf J}$ on the left-hand side is a reminder that the
averaging operation depends on the value of ${\bf J}$.

\item We define the averaged frequencies and forcing functions
\be
{\bar \omega}_\alpha({\bf J}) \equiv \langle
\omega_\alpha(\psi_\alpha,{\bf J}) \rangle_{\bf J}~,
\label{baromegadef}
\ee
and
\be
{\bar G}^{(1)}_\lambda({\bf J}) \equiv \langle
G^{(1)}_\lambda(\bfpsi,{\bf J}) \rangle_{\bf J}~.
\ee

\item We solve a set of ordinary differential equations in the slow
  time parameter
\be
{\tilde \lambda} = \epsilon \lambda~,
\ee
for two sets of auxiliary functions ${\bar \chi}_\alpha({\tilde \lambda})$ and
${\cal J}_\lambda({\tilde \lambda})$.  This set of ordinary differential equations is
\bes
\bea
\label{barchieqn}
\frac{d {\bar \chi}_\alpha}{d {\tilde \lambda}} &=& {\bar \omega}_\alpha(\bfcalJ({\tilde
  \lambda}))~,\\
\frac{d {\cal J}_\lambda}{d {\tilde \lambda}} &=& {\bar
G}^{(1)}_\lambda(\bfcalJ({\tilde  \lambda}))~.
\label{calJeqn}
\eea
\ees
Note that for this step, one does not need to specify a value of
$\epsilon$.

\item We can  then write down the adiabatic solutions:
\bes
\label{adiabaticsolutionapp}
\bea
\label{Jlambdaans}
J_\lambda(\lambda,\epsilon) &=& {\cal J}_\lambda(\epsilon \lambda), \\
\psi_\alpha(\lambda,\epsilon) &=& \Xi_\alpha\left[\frac{1}{\epsilon} {\bar \chi}_\alpha(\epsilon \lambda)~,
\bfcalJ(\epsilon \lambda) \right]~,
\label{psialphaans}
\eea
\ees
where the function $\Xi_\alpha(\chi,{\bf J})$ is defined implicitly by the equation
\be
\frac{\chi}{2 \pi} = \frac{\int_0^{\Xi_\alpha(\chi,{\bf J})}
  \frac{d\psi}{\omega_\alpha(\psi,{\bf J})}}{\int_0^{2\pi}
  \frac{d\psi}{\omega_\alpha(\psi,{\bf J})}}~.
\label{Xidef}
\ee
and satisfies
\be
\Xi_\alpha(\chi+2\pi,{\bf J}) = \Xi_\alpha(\chi,{\bf J}) + 2\pi~.
\label{periodic1}
\ee
\end{enumerate}

We now turn to the derivation of this result.  We start by rewriting the differential Eqs.~(\ref{gs}) in terms of the new variables
$(\chi_\alpha,J_\lambda)$, defined implicitly by the relation
\be
\psi_\alpha(\chi_\alpha,{\bf J}) \equiv \Xi_\alpha(\chi_\alpha,{\bf
  J})~.
\label{transf1}
\ee
All of the functions appearing in the differential equations are
expressed as functions of the new phases $\chi_\alpha$; they must be
periodic functions of each $\chi_\alpha$ by virtue of the property (\ref{periodic1}).
Using the definitions (\ref{Xidef}), (\ref{averagedef}) and
(\ref{baromegadef}) the result can be written in the
form
\bes
\label{gs1}
\bea
{\dot \chi}_\alpha &=& {\bar \omega}_{\alpha}({\bf J}) +
\epsilon \frac{{\bar \omega}_\alpha({\bf
    J})}{\omega_\alpha(\chi_\alpha,{\bf J})}
g_\alpha^{(1)}(\bfchi,{\bf J}) +
O(\epsilon^2)~,\nonumber\\
&&1 \le \alpha \le N~, \\
{\dot J}_\lambda &=& \epsilon G^{(1)}_\lambda(\bfchi, {\bf J}) +
\epsilon^2 G^{(2)}_\lambda(\bfchi, {\bf J}) + O(\epsilon^3),~
1 \le \lambda \le M~.\nonumber\\&&
\eea
\ees

This system of equations is now in a form to which the results of 
Ref.\ \cite{FH} can be applied; the variables $(\chi_\alpha,J_\lambda)$
are generalized action-angle variables.  The
averaging operation defined in \cite{FH}, a straightforward averaging
with respect to the phases $\chi_\alpha$, coincides with the
definition (\ref{averagedef}) used here, because of the definition
(\ref{transf1}).  The results of Ref.\ \cite{FH} now imply that the
leading order solution for $J_\lambda$ is of the form given by Eqs.\
(\ref{Jlambdaans}) and (\ref{calJeqn}).  They also imply that the
leading order solution for $\chi_\alpha$ is of the form
$\chi_\alpha(\lambda,\epsilon) = {\bar \chi}_\alpha(\epsilon
\lambda)/\epsilon$, where ${\bar \chi}_\alpha$ satisfies the
differential equation (\ref{barchieqn}).  Combining this with the
definition (\ref{transf1}) now yields the result (\ref{psialphaans}).

\section{Perturbation of Keplerian Orbits}
\label{KepOsc}
Here we derive the osculating element equations for a Keplerian orbit experiencing a force in the plane of the orbit, ${\bf f} = -\mu {\bf r}/r^3$. In this case, we can take the orbital plane to be the x-y plane. The orbit is described by four parameters --- the semimajor axis, $a$, the eccentricity, $e$, the argument of perihelion, $\omega$, and the time of pericenter passage, $T_0$. (The restriction to a plane gets rid of the other two orbital constants.) The orbit is elliptical and described by
\bea
r &=& \frac{a(1-e^2)}{1+e\cos(u-\omega)} = a\left(1-e\cos E\right)~,\\
\dot{u} &=& \sqrt{\frac{\mu}{a^3(1-e^2)^3}} \, \left(1+e\cos(u-\omega)\right)^2~,
\ena
in which $u$ is the argument. It is usual to call $v=u-\omega$ the {\it true anomaly} and $E$ defined by the first equation above is the {\it eccentric anomaly}. The time of pericenter passage is given implicitly by
\be
\int_0^{v_0} \frac{\rmd v'}{(1+e\cos v')^2} = \sqrt{\frac{\mu}{a^3(1-e^2)^3}} \left(t_0-T_0\right)~,
\en
where $v_0 = v(t_0)$.

Under the action of a force in the orbital plane with radial component $R'$ and tangential component $S'$, the Gaussian perturbation equations predict the following evolution equations for the four orbital elements \cite{bookPert}:
\bea
\dot{a} &=& \sqrt{\frac{a(1-e^2)}{\mu}} \frac{2a}{1-e^2} \left(e\sin v R' + \frac{p}{r} S'\right)~, \\
\dot{e} &=&  \sqrt{\frac{a(1-e^2)}{\mu}} \left[ \sin v R' + (\cos v + \cos E) S'\right]~,
\label{edot}\\
\dot{\omega} &=& \frac{1}{e} \sqrt{\frac{a(1-e^2)}{\mu}} \left[ -\cos v R' + \left(1+ \frac{r}{p} \right) \sin v S' \right]~, \nonumber\\&&\\
\dot{T_0}&=& -\frac{a^2(1-e^2)}{\mu e}\left[ \left(\cos v-2e \frac{r}{p} \right)R'\right. \nonumber\\
 &-&\left. \left(1+\frac{r}{p}\right) \sin v S'\right] -\frac{3}{2}\frac{\dot{a}}{a} (t-T_0)~.
\ena
If we consider the true anomaly, $v$, then since $v = u-\omega$, $\dot{v} = \dot{u} - \dot{\omega}$. By the definition of the osculating elements, the value of $\dot{u}$ is always given by the geodesic value, and so we see that the evolution of the true anomaly differs from integrating the instantaneous time-evolving geodesic equation by the $\dot{\omega}$ term. This can also be seen by differentiating the orbit equation and using that both $r$ and $\dot{r}$ are consistent with the instantaneous geodesic to obtain
\bea
\dot{v} &=& \frac{\sqrt{\mu a(1-e^2)}}{r^2}
+ \frac{\dot{e}}{e} \frac{\cos v}{\sin v} \nonumber\\
&-& \left(\frac{\dot{a}}{a} - \frac{2\dot{e}e}{1-e^2}\right)\left(\frac{1+e\cos v}{e\sin v}\right)~.
\label{KepTA}
\ena
The first term is the geodesic $\dot{v}$, while the other terms arise as a result of the perturbation. Although this equation looks singular at turning points, $\sin v = 0$, substitution of the expressions for $\dot{a}$, $\dot{e}$ and the geodesic equations gives the necessary calculations and the expression reduces to $-\dot{\omega}$, as it should.

\section{Drag force in Schwarzschild geometry}
\label{A:SchDrag}
In order to understand the effect that leads to an increase of eccentricity we can consider a Schwarzschild BH system, in which the same effect is seen, but which is easier to analyze and to understand. The osculating element equation for the evolution of the eccentricity, Eq.~(\ref{OscEcc}), in the case of a nonrotating BH reduces to Eq.~(37) in~\cite{pound08} and has the form
\be
\frac{de}{dv}  = \mathcal{R}(p, e, v) a^r + \mathcal{T}(p, e, v) a^{\phi}~.
\ee
We use a drag force to perturb the orbit which takes a very simple form $a^r = -\gamma u^r, \; a^{\phi} = -\gamma u^{\phi}$. The velocities, in Schwarzschild coordinates, are
\bea
u^r &=& e \sin{v} \sqrt{ \frac{p - 6 - 2e \cos{v}}{p(p-3-e^2)}}~,\\
u^{\phi} &=& \frac{(1+e\cos{v})^2} {pM\sqrt{p-3-e^2} }~.
\ena
The equation for $de/dv$ is integrable for this perturbing force if changes to $e$ and $p$ are ignored over the orbit and the result can expressed in terms of elliptic integrals. However, this is quite
messy and we are primarily interested in the leading order correction to the orbit. We make a weak field expansion ($M/p << 1$) of the terms entering this equation:
\bea
\mathcal{R} \approx \frac{p^{2}}{M} \left( \mathcal{R}_0(e,v) + \frac{M}{p}\mathcal{R}_1(e,v) + O(M^2/p^2) \right)~, \\
\mathcal{T} \approx {p^{3}}{M} \left( \mathcal{T}_0(e,v) + \frac{M}{p}\mathcal{T}_1(e,v) + O(M^2/p^2) \right)~,
\ena
here we do not go beyond the first correction to the Keplerian term. Similarly, we find for the velocities
\begin{align}
u^r &= \sqrt{\frac{M}{p}}   u^r_0(e,v) \left( 1 + \frac{M}{p} u^r_1(e,v) + O(M^2/p^2) \right)~,&\\
u^{\phi} &= \sqrt{\frac{M^3}{p^3}}   u^{\phi}_0(e,v) \left( 1 + \frac{M}{p} u^{\phi}_1(e,v) + O(M^2/p^2) \right)~.&
\end{align}
The explicit form of the terms in these expansions is
\begin{widetext}
\bea
u^r_0 &=& e \sin{v}~,
u^r_1 = \left(-\frac3{2} - e\cos{v} +\frac1{2}e^2\right)~,\\
u^{\phi}_0 &=&  (1+e\cos{v})^2, \; u^{\phi}_1 =  \left( \frac3{2} + \frac1{2}e^2\right)~,\\
\mathcal{R}_0 &=& \frac{\sin{v} }{(1+e\cos{v})^2},\; \mathcal{R}_1 = 3\mathcal{R}_0(1-e^2)~,\\
\mathcal{T}_0 &=& \frac{(e\cos{v}+2)\cos{v}+e}{(1+e\cos{v})^4}~, \\
\mathcal{T}_1 &=& \frac{2e + 6\cos{v}+2e\cos^2{v}-e^2\cos^3{v}-5e^2\cos{v}- 3e^3\cos^2{v}-e^3}
{(1+e\cos{v})^4}~.
\ena
\end{widetext}
The leading order terms give us the Newtonian perturbation of the eccentricity \eqref{edot} with
perturbing force components $R' = -\gamma \dot{r}$, $S' = -\gamma r\dot{\phi}$. Overall, the Newtonian term is
\bea
\left(\frac{de}{dv}\right)_{00} = -2\gamma  p^{3/2} \frac{e+\cos{v}}{(1+e\cos{v})^2}~.
\ena
This equation can be integrated over an orbit, keeping $e,p$ on the right-hand side constant, to give
\be
\delta e(v) = -2\gamma p^{3/2}\frac{\sin{v}}{1 + e \cos{v}}~.
\en
It is clear that in the Newtonian case there is no secular change in the eccentricity. Note also that the individual components (radial and azimuthal) of the perturbation are \emph{not} zero
after integration over one orbit, but they are exactly equal and opposite in sign. We now consider the first relativistic corrections. First, we note that the perturbations $\mathcal{R}_1$ and $u^{\phi}_1$ are independent of $v$, and so we can reabsorb these into a redefinition of $\gamma\rightarrow\gamma'$ where
$$
\gamma' =\left[ 1+ \frac1{2p}(3+e^2) \right] \left[ 1+ \frac3{p}(1-e^2) \right] \gamma~,
$$
and so the rescaled leading order term still averages to zero, as it is proportional to the Newtonian expression. There remain two perturbations, one that comes from the radial velocity perturbation, $\mathcal{R}_0 u^r_1$, and one that comes from the relativistic correction to the orbit's response to the azimuthal perturbation, $u^{\phi}_0 \mathcal{T}_1$. The velocity perturbation contributes
\bea
\left( \frac{de}{dv}\right)_{ 01} &=& -\gamma' \frac{p^{1/2}e\sin^2{v}}{(1+e\cos{v})^2} \nonumber\\
&\times& \left[\frac{1}{2}(-3-2e\cos{v}+e^2) -\frac1{2} (3+e^2)\right]\nonumber \\
&=& \gamma' p^{1/2} \frac{e\sin^2{v} (3+e\cos{v})}{ (1+e\cos{v})^2}~.
 \ena
Note that this term is always positive and so it will lead to an increase in the eccentricity. This can be interpreted as an additional radial force which acts at each point of the orbit in the direction of motion
slowing down the effective radial velocity in the force, which leads to the increase of eccentricity.

The second part of the perturbation, $u^{\phi}_0 \mathcal{T}_1$, contributes
\bea
\left( \frac{de}{dv}\right)_{10} &=& \gamma' ep^{1/2} \nonumber\\
&\times& \left[1-e^2 + \cos^2{v}(1+ e\cos{v}) -e(\cos{v}+e)\right] \nonumber\\
&\times&(1+e\cos{v})^{-2}~.
\ena
Note that
the last term is proportional to the Newtonian term and therefore averages to zero
so we can ignore this term. The remaining part is always positive and also drives an increase in eccentricity. This time the extra term can be interpreted as an additional azimuthal force which further boosts the effective azimuthal velocity in the force and once again leads to the increase in the eccentricity.

We note that both of these perturbations, and also the Keplerian term,  are proportional to eccentricity, and they will not drive a circular orbit to become eccentric. In fact, the contribution from the relativistic correction to the velocity is equal to that coming from the correction to the orbital response. Taking the difference,
\bea
\left( \frac{de}{dv}\right)_{01} - \left( \frac{de}{dv}\right)_{10} &=&
\gamma' \frac{ep^{1/2}}{(1+e\cos{v})^2} \nonumber \\
&\times& \left[e(\cos{v}+e) + 2\sin^2{v} \right.\nonumber\\
&&\left.- 2\cos^2{v}(1+e\cos{v})\right]~.\nonumber\\&&
\ena
The first term in the square bracket is proportional  to the Newtonian term and therefore vanishes after averaging. The remaining term can be integrated analytically,
 \be
 \int_0^v dv' \frac{\sin^2{v'} - \cos^2{v'}(1+e\cos{v'})}{(1+e\cos{v'})^2} = - \frac{\sin{v}\cos{v}}{1+e\cos{v}}.
 \en
which is also zero after integration over one orbit. We conclude that the leading order relativistic correction in the perturbation equation predicts the increase in eccentricity that we observe numerically. This secular change comes equally from the first order correction to the radial velocity and the first order  correction to the orbital response to an azimuthal perturbation. The relativistic corrections can be thought of as an extra force which slows down the effective radial motion and accelerates the effective azimuthal motion that enter the drag force. The radial drag force is correspondingly reduced, while the azimuthal drag force is increased and both drive a secular increase in eccentricity. The equality of the two parts of the force may reflect some hidden symmetry in the equations. The response of the orbit to a perturbation depends on the velocity at each point along the orbit, and we are using that same velocity to prescribe the perturbation in this case, which might explain why the net contribution from the two terms is equal. However, the osculating element equations are not explicit in how they depend on the instantaneous velocity, so this is only a speculation.

\bibliography{OscEltKerr}

\end{document}